\documentclass[reprint,superscriptaddress,amsmath,amsfonts,amssymb,aps,prl,showkeys]{revtex4-2}
\usepackage{hyperref}
\usepackage{graphicx}
\usepackage{braket}
\usepackage[dvipsnames]{xcolor}
\hypersetup{
  breaklinks = {true},
  citecolor = {blue},
  colorlinks = {true},
  linkcolor = {red},
}

\usepackage[T1]{fontenc}
\usepackage{soul}
\usepackage{physics}
\usepackage{mathtools}
\usepackage[dvipsnames]{xcolor}
\usepackage{braket} 
\usepackage{upgreek}
\usepackage{bm}

\usepackage{ulem}

\begin{document}

\title{Disorder-induced Delocalization in Magic-Angle Twisted Bilayer Graphene}

\author{Pedro Alc\'azar Guerrero}
\affiliation{Catalan Institute of Nanoscience and Nanotechnology (ICN2), CSIC and BIST, Campus UAB, Bellaterra, 08193 Barcelona, Spain}
\affiliation{Department of Physics, Universitat Autònoma de Barcelona (UAB), Campus UAB, Bellaterra, 08193 Barcelona, Spain}
\author{Viet-Hung Nguyen}
\affiliation{Institute of Condensed Matter and Nanosciences, Universit\'e catholique de Louvain (UCLouvain), Chemin des \'etoiles 8, B-1348 Louvain-La-Neuve, Belgium}
\author{Jorge Martínez Romeral}
\affiliation{Catalan Institute of Nanoscience and Nanotechnology (ICN2), CSIC and BIST, Campus UAB, Bellaterra, 08193 Barcelona, Spain}
\affiliation{Department of Physics, Universitat Autònoma de Barcelona (UAB), Campus UAB, Bellaterra, 08193 Barcelona, Spain}
\author{Aron W. Cummings}
\affiliation{Catalan Institute of Nanoscience and Nanotechnology (ICN2), CSIC and BIST, Campus UAB, Bellaterra, 08193 Barcelona, Spain}
\author{Jos\'e-Hugo Garcia}
\affiliation{Catalan Institute of Nanoscience and Nanotechnology (ICN2), CSIC and BIST, Campus UAB, Bellaterra, 08193 Barcelona, Spain}
\author{Jean-Christophe Charlier}
\affiliation{Institute of Condensed Matter and Nanosciences, Universit\'e catholique de Louvain (UCLouvain), Chemin des \'etoiles 8, B-1348 Louvain-La-Neuve, Belgium}
\author{Stephan Roche}
\affiliation{Catalan Institute of Nanoscience and Nanotechnology (ICN2), CSIC and BIST,
Campus UAB, Bellaterra, 08193 Barcelona, Spain}
\affiliation{ICREA--Instituci\'o Catalana de Recerca i Estudis Avan\c{c}ats, 08010 Barcelona, Spain}

\begin{abstract}
Flat bands in moiré systems are exciting new playgrounds for the generation and study of exotic many-body physics phenomena in low-dimensional materials.
Such physics is attributed to the vanishing kinetic energy and strong spatial localization of the flat-band states.
Here we use numerical simulations to examine the electronic transport properties of such flat bands in magic-angle twisted bilayer graphene in the presence of disorder.
We find that while a conventional downscaling of the mean free path with increasing disorder strength occurs at higher energies, in the flat bands the mean free path can actually increase with increasing disorder strength.
 {This phenomenon is also captured by the disorder-dependent quantum metric, which is directly linked to the ground state localization.}
This disorder-induced delocalization suggests that weak disorder may have a strong impact on the exotic physics of magic-angle bilayer graphene and other related moiré systems.

\end{abstract}

\maketitle

\textit{Introduction}. Moiré systems, such as magic-angle twisted bilayer graphene (MATBLG), have proven to be ideal structures for generating flat bands and strongly correlated physics \cite{Cao2018, Cao2018b, Stepanov2020, pnas1108174108, Andrei2021}. The unexpected and spectacular observation of superconductivity and Mott insulating phases in MATBLG has been attributed to the vanishing kinetic energy and real-space state localization \cite{Tilak2021}. This triggers a dominant contribution of the Coulomb interaction, which generates a wealth of emerging many-body physics phenomena \cite{Cao2018, Cao2018b, Stepanov2020, pnas1108174108, Andrei2021, Tilak2021, Serlin2020, Saito2020, Balents2020, PhysRevLett.127.197701, Xie2021, Pierce2021, PhysRevLett.126.137601, Park2022, Jaoui2022, Paul2022, Klein2023, Tian2023}. Other studies have probed the robustness of this behavior and its relationship with the emerging superlattice potential, which surprisingly occurs for periodic as well as aperiodic (such as quasicrystalline order) systems \cite{Ahn2018, Uri2023, lai2023imaging}.

The role of the moiré superlattice potential in producing such “anomalous electronic spatial localization” and vanishing velocity of the flat bands is now well documented \cite{PhysRevResearch.2.023237, LEDWITH2021168646, PhysRevB.105.224508}. However to date, relatively little is known about the impact of superimposed structural (i.e. twist angle)\cite{ciepielewski2024transport} and electrostatic disorder, which are ubiquitous in real samples and likely play a role in the stability of the flat band-induced localization and the resilience of strong correlation effects. In particular, the impact of disorder on the transport and localization properties remains to be examined. It has been shown that structural relaxation (which will be substrate-dependent) leads to a renormalized dispersion of the flat bands which may then impact the transport properties \cite{PhysRevB.99.205134, Hung2021, PhysRevLett.127.126405}. Additionally, in the context of general flat-band physics, several theoretical studies suggest that disorder could produce exotic phenomena such as inverse Anderson localization (observed experimentally in ultracold atoms) \cite{PhysRevLett.129.220403}, super-metallicity \cite{PhysRevB.103.075415, PhysRevB.108.155108}, or deviation from the usual scaling theory of localization \cite{PhysRevB.88.224203}. {It is thus important to study the impact of disorder in such systems, as some amount of disorder is present in all fabricated samples \cite{PhysRevResearch.2.023325, Kazmierczak2021ER, deJong2022}.}

{In the last decade, topological materials have been widely studied due to their nontrivial ground state \cite{Thouless1982Aug, Victor1984Mar, Kane2005Sep, Haldane1988Oct}. A particularly fundamental quantity is the quantum geometric tensor, whose imaginary part is the Berry curvature, $\Omega_{ij}$, which drives most studied topological effects, while its real part ($g_{ij}$) is known as the quantum metric (QM) \cite{Provost1980Sep}. Recently $g_{ij}$ has been gaining much attention due to its emergence in linear \cite{TormaEssayQM, Queiroz1, Queiroz2, orbitalsusceptivilitywithoutberrycurvature, Kruchkov2023Dec} and nonlinear response theory \cite{Gao2023Jun, Gao2019Jun, Wang2021Dec, Das2023Nov}, although a direct link with an observable remains challenging \cite{TormaEssayQM}. In various flat-band systems, the QM seems to play an important role \cite{Kruchkov2023Jun, Mera2022Oct, Rhim2020Aug}, especially in the interacting phase \cite{Peotta2015Nov, Julku2020Feb, Peri2021Jan, Rossi2021, Torma2022}. In addition, the QM is at the heart of the modern theory of insulators \cite{theoryofinsulatingKohn, ModerntheoryinsulatingResta}, where it connects to the real part of the dipole-dipole fluctuations of the ground state, $\Re\left<\hat{r}_{i}\hat{r}_{j}\right>_c=g_{ij}$, indicating if the ground state is metallic (when the fluctuations diverge in the thermodynamic limit), or an insulator. Therefore $g_{ij}$ is directly linked to the electronic localization of the ground state \cite{MaximallylocalizedWannierfunctions, Localtheoryofinsulatingstate}. However, despite recent advances, the behaviour of the QM in the presence of disorder is still an open question, particularly for disordered MATBLG.}

In this Letter, we use numerical simulations to investigate the impact of disorder on electronic transport in MATBLG. We implement a realistic tight-binding Hamiltonian which well captures the weakly dispersive flat bands in this system. With a rotation angle of ${\sim}1.1^\circ$ between the two layers, the moiré superlattice has a long range of periodicity, with more than {11,000} atoms in the unit cell. To account for the disorder-induced breaking of translational invariance, we employ a real-space linear-scaling quantum transport methodology that enables the simulation of disordered MATBLG systems containing several million atoms, thus allowing us to reach transport length scales that are relevant to experiments \cite{FAN20211}. {In addition, we compute the dimensionless quantum metric (from now on just the quantum metric), ${\cal G}_{ij}=g_{ij}/V$, which gives us a direct measure of the electronic localization of the ground state, via the Souza-Wilkens-Martin (SWM) rule \cite{SWMrule} and the polynomial expansion of the optical conductivity \cite{FAN20211, opticalconductivity}.} Our findings reveal a nontrivial evolution of transport characteristics for disorder strengths that do not fully suppress the flat bands. The combination of disorder-induced broadening and scattering leads to delocalization of the flat-band states, which manifests in an increase of the mean free path with increasing disorder. This is opposite to the more conventional behavior observed at higher energies away from the moiré flat bands, where stronger disorder reduces the mean free path. Such unconventional behavior disappears for disorder strong enough to break the intrinsic moiré-induced localization at the magic angle. {We have seen emergence of this exotic behavior in both the quantum transport and QM calculations.}

These findings suggest that weak disorder in MATBLG, by driving delocalization of the flat-band states, may have a significant impact on the relative strength of the Coulomb interaction at the heart of the exotic physics in this material. In this context, a metric for characterizing the delocalization of flat-band states may serve as an important quantity for characterizing the stability of exotic phases in such systems. {In addition we expect this result to be general for other systems with flat bands.}

\textit{Structural and tight-binding Hamiltonian models}. To build realistic magic-angle twisted graphene superlattices, we use molecular dynamics simulations with classical potentials to relax the structures \cite{Hung2021, Nguyen_2022, Lindsay2010, Kolmogorov2005, Leven2016}. We start with uniform Bernal-stacked bilayer graphene, twisted to an angle of ${\sim}1.1^\circ$, and optimize the structure until all the force components are smaller than 0.5 meV/atom. Intralayer forces are computed using the optimized Tersoff and Brenner potentials \cite{Lindsay2010}, whereas interlayer forces are modeled using the Kolmogorov–Crespi potentials \cite{Kolmogorov2005, Leven2016}.
The electronic properties of the twisted graphene systems are then computed using the $p_z$ tight-binding (TB) Hamiltonian,
\begin{equation}
\mathcal{\hat{H}} = \sum_{n} \varepsilon_n \ket{\phi_n} \bra{\phi_n} + \sum_{n,m} t (\vec{r}_{nm}) \ket{\phi_n} \bra{\phi_m},
\end{equation}
where $\ket{\phi_n}$ describes the $p_z$ orbital on carbon site $n$ with position $\vec{r}_{n}$, $\varepsilon_n$ is the electrostatic potential at carbon site $n$, and $\vec{r}_{nm} = \vec{r}_{m} - \vec{r}_{n}$. The hopping energies $t (\vec{r}_{nm})$ between carbon sites are given by the standard Slater–Koster expression \cite{Trambly2010,Trambly2012,Hung2021,Nguyen_2022}
\begin{align}
t ({\vec r}_{nm}) &= \cos^2 (\phi_{nm}) V_{pp\sigma} (r_{nm}) \nonumber \\
&+ \sin^2 (\phi_{nm}) V_{pp\pi} (r_{nm}),
\end{align}
where the direction cosine of $\vec{r}_{nm}$ along the $z$-axis is $\cos(\phi_{nm}) = z_{nm}/r_{nm}$. The distance-dependent Slater-Koster parameters are \cite{Trambly2012}
\begin{align}
V_{pp\pi} (r_{nm}) &= V_{pp\pi}^0 \exp \left[ q_\pi \left( 1 - \frac{r_{nm}}{a_0}  \right) \right] F_c (r_{nm}), \nonumber \\
V_{pp\sigma} (r_{nm}) &= V_{pp\sigma}^0 \exp \left[ q_\sigma \left( 1 - \frac{r_{nm}}{d_0}  \right) \right] F_c (r_{nm}),
\end{align}
with a smooth cutoff function $F_c (r_{nm}) = \left[ 1 + \exp \left(( r_{nm}-r_c)/\lambda_c\right) \right]^{-1}$.
To model the flat electronic bands of relaxed TBLG at the magic angle $\sim$$1.1^\circ$, the TB parameters are adjusted to $V_{pp\pi}^0 = - \gamma_{0}= -2.7 \,~\mathrm{eV}$, $V_{pp\sigma}^0 = 367.5 \,~\mathrm{meV}$, $q_\pi/a_0 = q_\sigma/d_0 = 22.18 \,~\mathrm{nm}^{-1}$, $a_0 = 0.1439 \,~\mathrm{nm}$, $d_0 = 0.33 \,~\mathrm{nm}$, $r_c = 0.614 \,~\mathrm{nm}$, and $\lambda_c = 0.0265 \,~\mathrm{nm}$ \cite{Nguyen_2022}.
Finally, disorder is introduced via an Anderson potential by assigning random values to the onsite energies within a uniform distribution with interval $\varepsilon_n \in [-W/2,W/2]$. {We consider values of $W$ from $\gamma_0/6$ to $2\gamma_0$, corresponding to bulk mean free paths from a few tens to a few hundred nanometers \cite{FoaTorres}, covering the range typically measured in graphene on SiO$_2$ \cite{Tan2007}.}

\textit{Quantum transport methodology}. To investigate electronic transport in the relaxed twisted bilayer graphene structure, we use the linear-scaling quantum transport methodology detailed in Ref.\ \citenum{FAN20211}. Specifically, we calculate the mean square displacement of an initial electronic state $\ket{\psi(0)}$ as a function of Fermi energy and time,
\begin{equation}
\Delta X^2(E,t) = \frac{ \langle\psi_X(t) | \delta(E-\mathcal{\hat H}) | \psi_X(t)\rangle }{ \rho(E) },
\end{equation}
where $\ket{\psi_X(t)} = [ \hat{X},\hat{U}(t) ] \ket{\psi(0)}$, $\hat{X}$ is the position operator, $\hat{U}(t) = \exp(-\text{i}\mathcal{\hat  H}t/\hbar)$ is the time evolution operator, and $\rho(E) = \braket{\psi(0) | \delta(E-\mathcal{\hat H}) | \psi(0)}$ is the density of states. The time evolution operator and the energy projection operator $\delta(E-\mathcal{\hat  H})$ are both expanded in a numerically efficient way using Chebyshev polynomials \cite{FAN20211, WeisseKPM}. Here we use 3500 polynomials, corresponding to a Gaussian energy broadening of 23 meV. We use a timestep of $10$ fs for
{transport in the clean system and $1$ fs for the disordered cases.}
The initial state $\ket{\psi(0)}$ is chosen to be a random-phase state, {which is a standard approach for the efficient calculation of material properties over the entire Hamiltonian spectrum; see Refs.\ [\citenum{FAN20211}] and [\citenum{WeisseKPM}] for more details.}

From the mean square displacement we calculate the time-dependent diffusion coefficient $D(E,t)$ and extract the mean free path $\ell(E)$ from its saturated value at long times, $D_\text{max}(E)$, according to
\begin{align}
D(E,t) &= \frac{1}{2}\frac{\text{d}}{\text{d}t}\Delta X^2(E,t), \\
\ell(E) &= \frac{2D_\text{max}(E)}{v(E)},
\end{align}
where $v(E)$ is the Fermi velocity of disorder-free MATBLG. In our transport simulations, a system with a $16 \times 16$ tiling of the MATBLG unit cell is considered, containing about 2.9 million atoms.

\begin{figure}[tbh]
\includegraphics[width=\columnwidth]{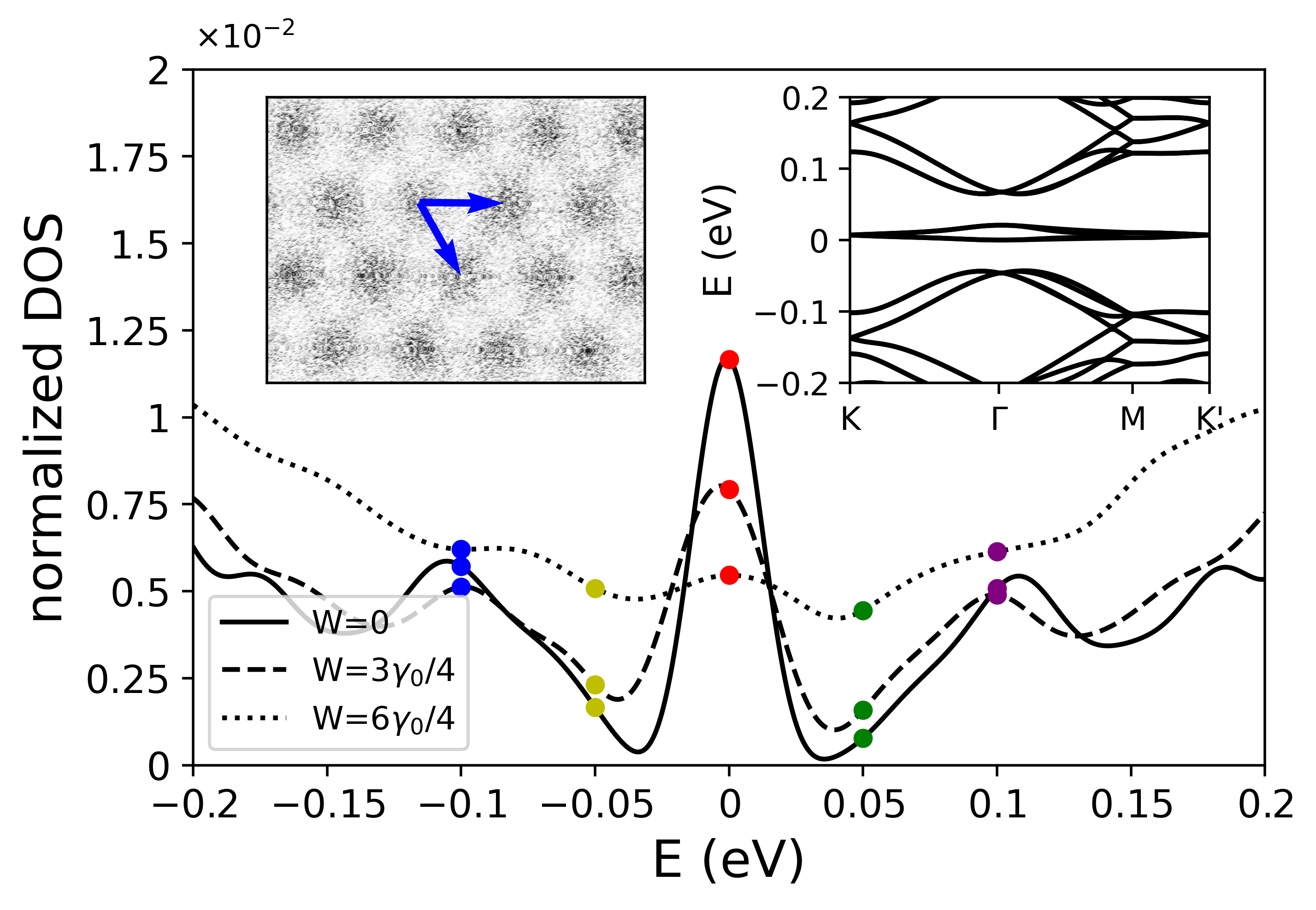}
\caption{Total density of states of MATBLG for disorder strengths of $W={3\gamma_0/2}$ (dotted line), $W=3\gamma_0/4$ (dashed line), and no disorder (solid line). Left inset: local density of states (LDoS) of the clean system at charge neutrality where higher LDoS (darker zones) is concentrated in the AA moiré regions. The blue arrows indicate the moiré unit cell lattice vectors. Right inset: bandstructure of MATBLG, with the flat bands clearly present around $E=0$. }
\label{fig1}
\end{figure}

{\textit{Quantum metric}. To compute the quantum metric we make use of the SWM rule \cite{SWMrule}, which relates it to the optical conductivity (via the fluctuation-dissipation theorem) as
\begin{equation}\label{eq:SWMrule}
    \int_0^{\infty}\dfrac{d\omega}{\omega}\Re{\sigma_{xx}(\omega)}=\dfrac{\pi e^2}{h}{\cal G}_{xx},
\end{equation}
where ${\cal G}_{xx}$ is the dimensionless quantum metric, $e$ is the electric charge, $h$ is the Planck constant, $\sigma(\omega)$ is the optical conductivity, and $\omega$ is the frequency. This QM is linked to the invariant part of the spread of the Wannier functions \cite{MaximallylocalizedWannierfunctions},
$\Omega_{\mathrm{I}} = \mathrm{Tr}\{{\cal G}_{xx}\} \cdot V/(2\pi)^2$,
where the trace is over the Cartesian indices. Here, a small QM relates to a small Wannier spread, and thus to a strongly localized ground state. On the other hand, when the QM increases continuously with system size, it indicates delocalization of the Wannier function and the ground state. Here we compute $\sigma(\omega)$ from the Kubo formula \cite{Kubo}, making use of a Chebyshev polynomial expansion \cite{opticalconductivity} with a broadening of 66 meV and obtain convergence of ${\cal G}_{xx}$ with system size and number of polynomials \cite{suppmaterial}. }

\textit{Electronic properties of disordered MATBLG}. We first start by analysing the impact of disorder on the electronic structure of MATBLG through its impact on the total (DoS) and local density of states (LDoS). In Fig.\ \ref{fig1} we plot the DoS for Anderson disorder strengths of $W = 0$, $3\gamma_0/4$, and ${3\gamma_0/2}$. For reference, we show the band structure of the clean case in the right inset and the LDoS at charge neutrality ($E = 0$) in the left inset. The flat bands and corresponding localization of states in AA regions are well reproduced in the absence of disorder. In the main panel, the presence of a strong peak in the DoS at $E=0$ highlights the presence of the moiré-induced flat bands. The role of disorder is to broaden and reduce this peak, which remains clearly visible at $W=3\gamma_0/4$ before finally being washed out for $W\geq{3\gamma_0/2}$, coinciding with the disappearance of AA spatial localization (see the right inset Fig.\ \ref{fig4} for $W=2\gamma_{0}$).

\begin{figure}[tbh]
\includegraphics[width=\columnwidth]{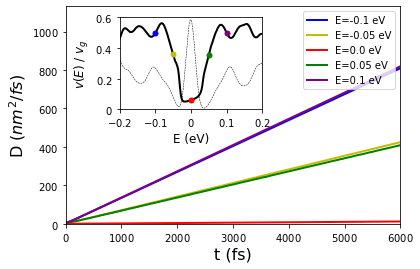}
\caption{Diffusion coefficient as a function of time for clean MATBLG at energies of $E = -100$ meV (blue), $-50$ meV (yellow), charge neutrality (red), $50$ meV (green), and $100$ meV (purple). Inset: the corresponding energy-dependent Fermi velocity (solid line), superimposed with the DOS of the clean case (dotted line, rescaled for clarity). }
\label{fig2}
\end{figure}

\textit{Quantum transport in clean and disordered MATBLG.} In Fig.\ \ref{fig2} we plot the time evolution of the diffusion coefficient of clean MATBLG. In  absence of disorder, transport is ballistic and the diffusion coefficient increases linearly with time, $D(E,t)=v^2(E) t$. This behavior is seen at all energies (different colored curves) in Fig.\ \ref{fig2}. From the slope of these curves we then extract the Fermi velocity of the MATBLG system, which we plot in the inset, relative to the velocity of single-layer graphene $v_\text{g}$.  {Note that while the Fermi velocity may not be uniform around the Fermi surface, here we are plotting its average over the Fermi surface at the indicated energies $E$.} Here we see that around charge neutrality the Fermi velocity is very low, $v < 0.1 v_\text{g}$, characteristic of the flat nature of the moiré bands.

\begin{figure}[tbh]
\includegraphics[width=\columnwidth]{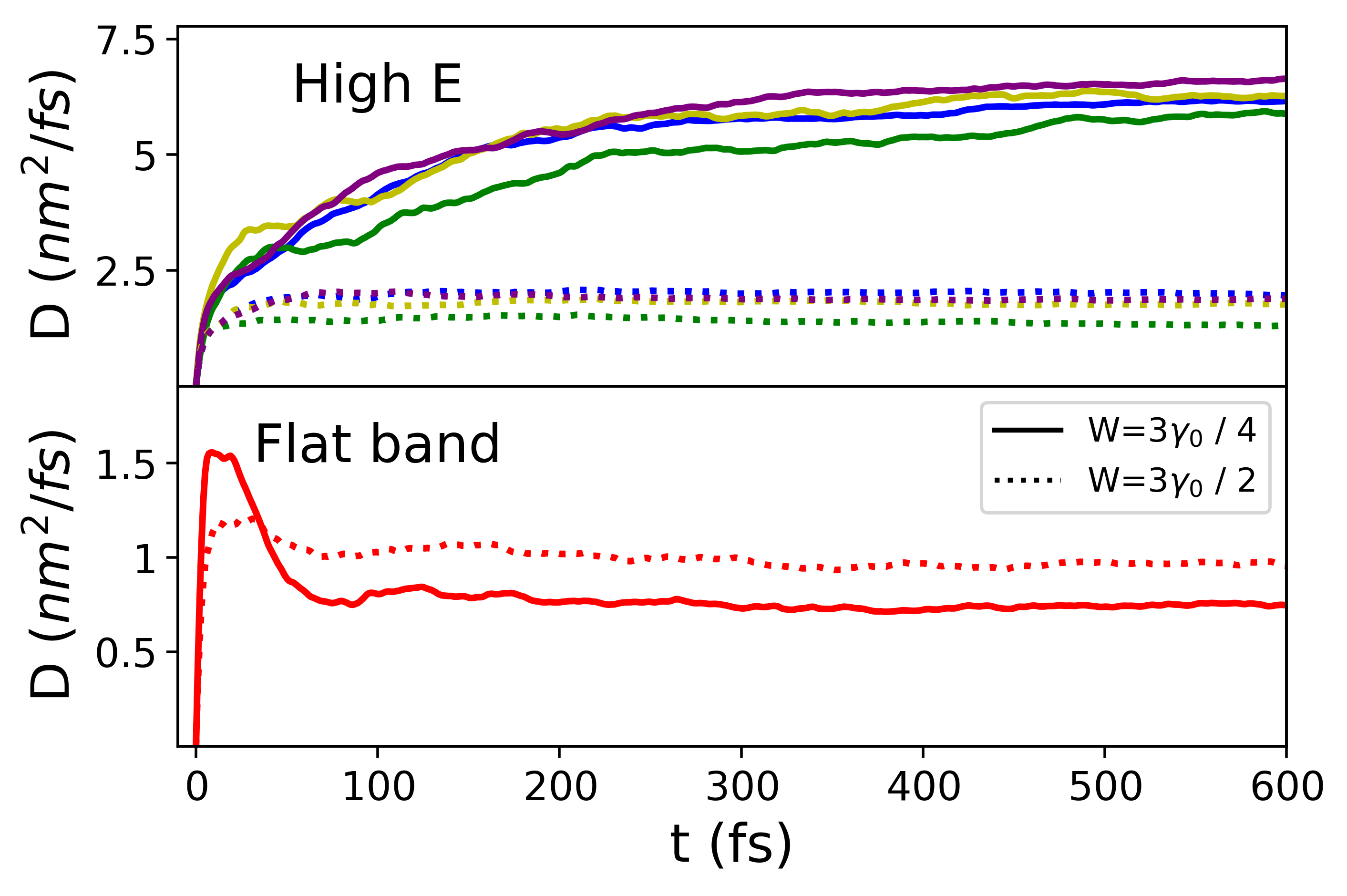}
\caption{{Time dependent diffusion coefficient at energies of $E = -100$ meV (blue), $-50$ meV (yellow), charge neutrality (red), $50$ meV (green), and $100$ meV (purple) for MATBLG under Anderson disorder strengths of $W=3\gamma_0/4$ (dashed) and $W={3\gamma_0/2}$ (dotted).} }
\label{fig3}
\end{figure}

\begin{figure}[tbh]
\includegraphics[width=\columnwidth]{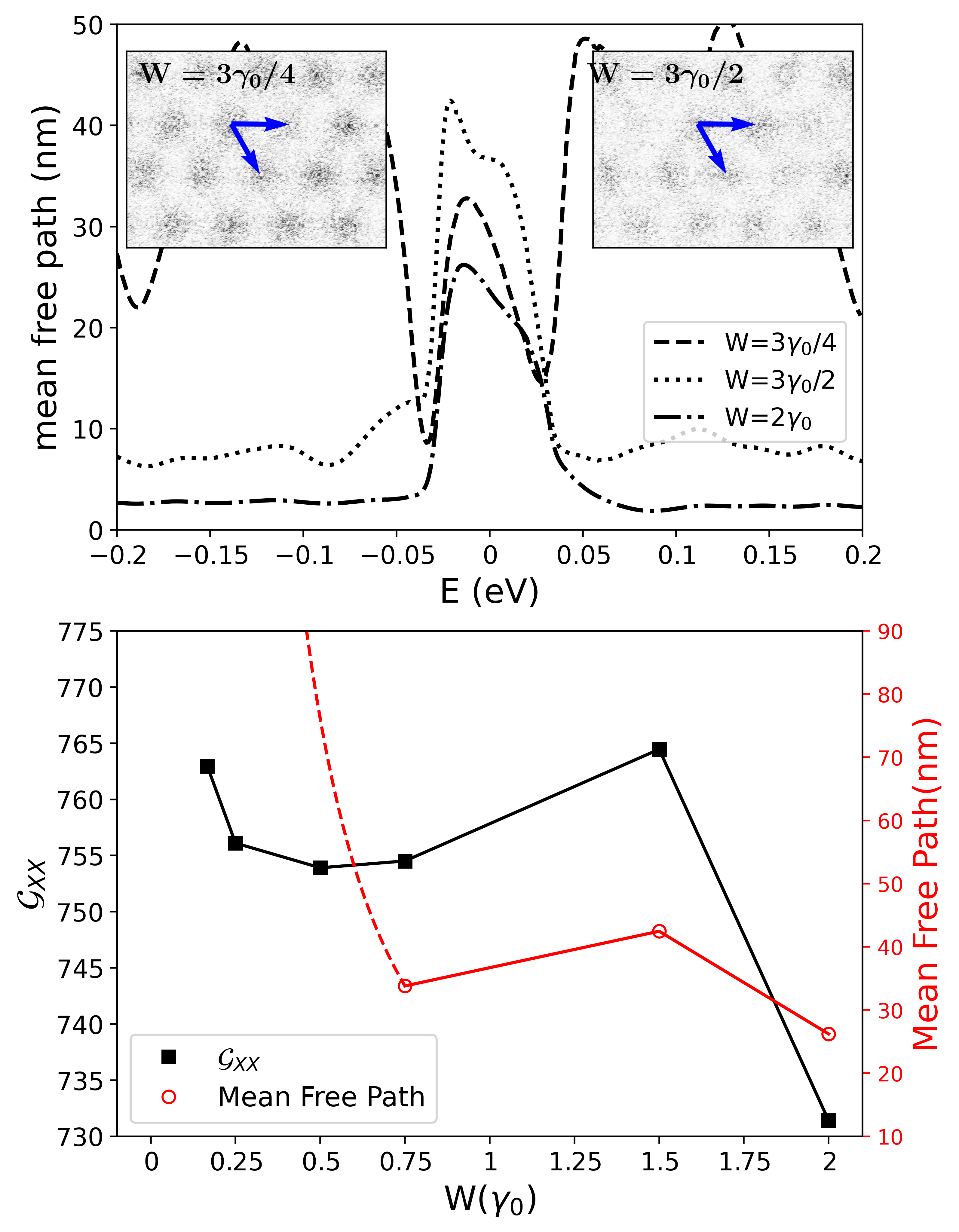}
\caption{
  {
    Top panel: mean free path for Anderson strengths of  $W=3\gamma_0/4$ (dashed line), {$W=3\gamma_0/2$} (dotted line) and $W=2\gamma_0$ (dash-dotted line). The insets show the LDoS at charge neutrality with disorders of  $W=3\gamma_0/4$ (left) and {$W=3\gamma_0/2$} (right). Bottom panel: quantum metric ${\cal G}_{xx}$ and mean free path at the flat band for different disorder strengths. {The red dashed line represent expected mean free path computed using the Fermi Golden rule for smaller disorders.}
  }
    }
\label{fig4}
\end{figure}

Next, in Fig.\ \ref{fig3} we examine electronic transport in disordered MATBLG. In the presence of Anderson disorder, the diffusion coefficients now saturate at long times for all energies. However, we observe a qualitative difference between transport within the flat bands compared to that at higher energies. When increasing disorder strength from  $W = 3\gamma_0/4 \rightarrow {3\gamma_0/2}$, $D$ decreases by a factor of $\sim$$4$ for all energies except at charge neutrality (red curve), where $D$ actually increases, opposite to typical behavior. This is illustrated further in Fig.\ \ref{fig4}, where we plot the mean free path $\ell(E)$ for three different disorder strengths. In the energy range corresponding to the moiré flat bands, for weaker disorder we see a clear increase of the mean free path with increasing disorder strength, opposite to the scaling behavior at higher energies. This increase in $\ell(E)$ actually coincides with a delocalization of the LDoS around charge neutrality in the presence of disorder, as highlighted in the left and right insets of Fig.\ \ref{fig4}. Here we note a slight electron-hole asymmetry in the mean free path, arising from an asymmetry in the band structure (inset of Fig.\ \ref{fig1}) and correspondingly in the Fermi velocity (inset of Fig.\ \ref{fig2}). A similar behavior of the electron-phonon coupling in MATBLG has been reported in Refs. \cite{Choi2018,Gadelha2022}.

We remark that stronger disorder will eventually suppress any remnant of the flat bands and thus reduce the mean free path following the scaling behavior $\ell\sim (\gamma_{0}/W)^{2}a_{cc}$, where $a_{cc}$ is the carbon-carbon spacing. This is seen when increasing disorder from $W={3\gamma_0/2}$ to $W=2\gamma_{0}$ in Fig.\ \ref{fig4}.

Therefore, the observed anomalous “disorder-induced delocalization” exists over a finite range of disorder strengths, and is maintained when disorder is low enough to preserve the moiré-induced flat-band states. This effect is driven by the disorder-induced broadening of the flat bands and the corresponding delocalization of states in real space. Using a simple argument based on the Fermi golden rule, the increase in the mean free path is driven by the reduction of the DoS and the corresponding scattering rate. For weaker disorder, following the scaling theory of localization, one expects that near the flat bands the localization length (related to the mean free path through the Thouless relationship) will reach values on the order of 100 nanometers for a disorder strength corresponding to the effect of electron-hole puddles generated by a silicon oxide substrate \cite{VanTuan2016}. Finally, we note that in the strong Anderson disorder limit, the mean free paths in disordered MATBLG are similar to those found in disordered monolayer graphene \cite{Lherbier2008}.

{\textit{Quantum metric in disordered MATBLG.} Figure \ref{fig4} (bottom panel) shows ${\cal G}_{xx}$ for different disorder strengths, and we compare its disorder-dependent evolution with the mean free path. Interestingly, the evolution between $W=0.75\gamma_{0}$ and $W=2\gamma_{0}$ is qualitatively similar for ${\cal G}_{xx}$ and $\ell$, indicating a disorder-induced delocalization mechanism. The increase of the QM for lower disorder is expected for weakly disordered cases, since the cleaner the system is the longer the corresponding mean free path and localization length. Note that the precise value of the mean free path is not shown for lowest disorders due to computational difficulty accessing the diffusive regime in the simulations}.
{However, we can make a rough estimate by using the Fermi golden rule when disorder only weakly affects the band structure, for which $l \propto 1/W^2$. Taking the smallest value of disorder with a converged mfp ($W = 3/4$), for which $l = 33.8$ nm, we extrapolate $l \approx 76$ nm for $W = 1/2$ and $l \approx 306$ nm for $W = 1/4$.}
{Here it is important to remark that such disorder-induced changes in the integrated quantum metric could also have an impact on MATBLG superconductivity, following some recent works relating the contribution of the integrated quantum metric to the superfluid weight and the critical temperature \cite{Julku2020Feb, Torma2022}}.



\textit{Conclusion and perspective}. Our findings show how disorder can interfere with the inherent superlattice-driven localization effect in MATBLG, giving rise to a nontrivial evolution of the transport length scales, {consistent with the evolution of the quantum metric}, as a function of disorder and energy. Although our model neglects the Coulomb interaction at the origin of the exotic many-body physics of MATBLG, understanding the impact of disorder on single-electron properties in flat-band systems could prove relevant to gauge the stability of exotic interaction-driven physics. Indeed, disorder-induced broadening of flat bands, {as seen in the DOS of Fig.\ \ref{fig1},} should result in a ``softening" of the Coulomb interactions, such that a possible regime of noninteracting electrons might be dominant for the considered Anderson disorder strengths, with Coulomb interactions playing a perturbative role driven by screening effects.

{On the other hand, we predict an increment of the quantum metric with the mean free path, which could translate to an extension of the wave packet and enhance the geometrical contribution to the superconductivity \cite{Torma2022, Julku2020Feb}. So we would expect the geometrical part of the superconductivity to gain importance for a certain range of disorder strength, as well as see a competition between these two contributions in stabilizing the superconducting state.}
This behavior could be sample-dependent and modulated by effects seen in most fabricated samples, such as substrate-induced corrugation, edge roughness, or twist angle disorder \cite{PhysRevResearch.2.023325, Kazmierczak2021ER, deJong2022}. Thus, the quantification of disorder-induced transport characteristics shown here should serve as an interesting metric to explore the stability and nature of the exotic transport phases reported in magic-angle twisted layered materials. {Moreover, the work could be extended applying this methodology to study the Coulomb interaction in such disordered systems (at least in the mean field).}

{Given the generality of the competition between localization due to geometry and delocalization induced by disorder, we expect these results to also be relevant to other types of flat-band systems in moiré structures or otherwise.}

\begin{acknowledgments}
P.A.G and J.M.R acknowledge T. Galvani for sharing optical conductivity codes. P.A.G., A.W.C., J.H.G and S.R acknowledge grant PCI2021-122035-2A-2 funded by MCIN/AEI/10.13039/501100011033 and European Union ``NextGenerationEU/PRTR” and the support from Departament de Recerca i Universitats de la Generalitat de Catalunya. J.H.G. acknowledge funding from the European Union (ERC, AI4SPIN, 101078370).  ICN2 is funded by the CERCA Programme/Generalitat de Catalunya and supported by the Severo Ochoa Centres of Excellence programme, Grant CEX2021-001214-S, funded by MCIN/AEI/10.13039.501100011033. 
V.-H.N. and J.-C.C. acknowledge financial support from the European Union’s Horizon 2020 Research Project and Innovation Program — Graphene Flagship Core3 (N$^{\circ}$ 881603), from the Flag-Era JTC projects “TATTOOS” (N$^{\circ}$ R.8010.19) and “MINERVA” (N$^{\circ}$ R.8006.21), from the Pathfinder project “FLATS” (N$^{\circ}$ 101099139), from the F\'ed\'eration Wallonie-Bruxelles through the ARC project “DREAMS” (N$^{\circ}$ 21/26-116), from the EOS project “CONNECT” (N$^{\circ}$ 40007563) and from the Belgium F.R.S.-FNRS through the research project (N$^{\circ}$ T.029.22F). Computational resources have been provided by the CISM supercomputing facilities of UCLouvain and the C\'ECI consortium funded by F.R.S.-FNRS of Belgium (N$^{\circ}$ 2.5020.11). Simulations were performed at the Center
for Nanoscale Materials, a U.S.\ Department of Energy Office of Science User Facility, supported by the U.S.\ DOE, Office of Basic Energy Sciences, under Contract No.\ 83336.
We acknowledge grant PID2022-138283NB-I00 funded by MICIU/AEI/ 10.13039/501100011033 and by “ERDF/EU”.
\end{acknowledgments}

\clearpage
\onecolumngrid
\section*{Suplementary Material}
 \vspace{.5cm}
\twocolumngrid
\section{SM I: Convergence of the Quantum metric with the number of Chebyshev polynomials and system size}

As stated in the main text, the quantum metric is linked to the localization of the ground state, via the modern theory of the insulating state \cite{ModerntheoryinsulatingResta}. In this theory, every ground state is assigned a localization length $\xi(V)$ that depends on the system size, $V$. If the localization length diverges in the thermodynamic limit the system is metallic, otherwise it will be an insulator. Disordered 2D systems are localized for every energy \cite{Lherbier2008}, and consequently $\xi$ will be finite. Because the quantum metric $\mathcal{G}_{ii}$ is linked to the localization of the system via the invariant part of the Wannier spread, $\Omega_{\mathrm{I}}$ \cite{haldanevanderblitlocalization}, we expect it to be scale quadratically with the localization length, $\mathcal{G}_{ii} \propto \xi^2$. Therefore $\mathcal{G}_{ii}$ will diverge within the thermodynamic limit for metals and be finite for insulators.

When dealing with a Chebyshev polynomial expansion of the Kubo formula for conductivity, we have an additional scaling related to the number of polynomials in the expansion, $M$. At this finite number of momenta the expansion becomes an $M$th-order polynomial of the Hamiltonian. This sets a real space cutoff which is of high relevance in our calculations, as happens with the so-called topological markers \cite{computationtopomarker}. In practice, we have to make sure that our results are converging both with system size and with the number of  Chebychev moments.

\begin{figure}[tbh]
\includegraphics[width=0.99\columnwidth]{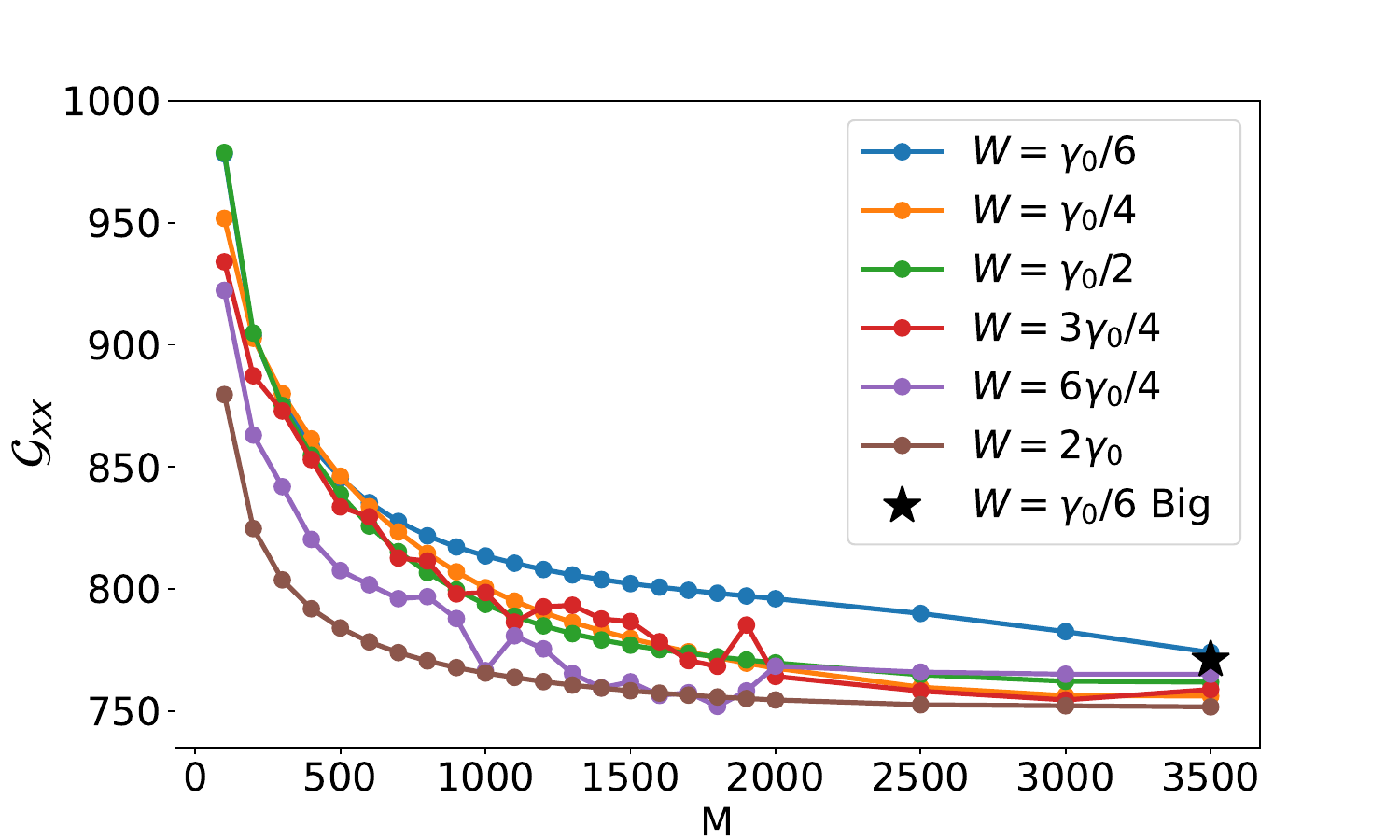}
\caption{$\mathcal{G}_{xx}$ as a function of the number of moments used in the Chebychev expansion for different disordered configurations. The black star marker shows $\mathcal{G}_{xx}$ for the lowest disorder at $M=3500$ in a system that is $1.5\times$ larger, indicating convergence with system size.}
\label{fig_convergence_with_moments}
\end{figure}

This can be seen in Fig.\ \ref{fig_convergence_with_moments}, where we have computed the unitless quantum metric presented in the main text, $\mathcal{G}_{xx}$, against the number of moments used in the expansion. As can be seen for all the disorders except $W = 1/6 \gamma_0$, the quantum metric has converged when reaching $M\sim2000$ moments. For the lowest disorder case, we expect it to take more moments to converge as the system is less localized and $\xi$ is larger. This is seen as the results for $W = 1/6\gamma_0$ start converging when reaching $M=3500$.

To see if we have reached the convergence in the system size, we have computed the quantum metric for $M=3500$ in the lowest disorder configuration for a system $1.5$ times bigger than our usual system. Reaching size convergence in this case ensures that the rest of the disorder strengths have also reached it, as it is the configuration with highest $\xi$. This can be seen in Fig.\ \ref{fig_convergence_with_moments}, where the black star marker indicates that the calculated quantum metric does not change when increasing the system size (compare to the orange results).

\section{SM II: Short-time evolution of the wave packet}

\begin{figure}[tbh]
\includegraphics[width=\columnwidth]{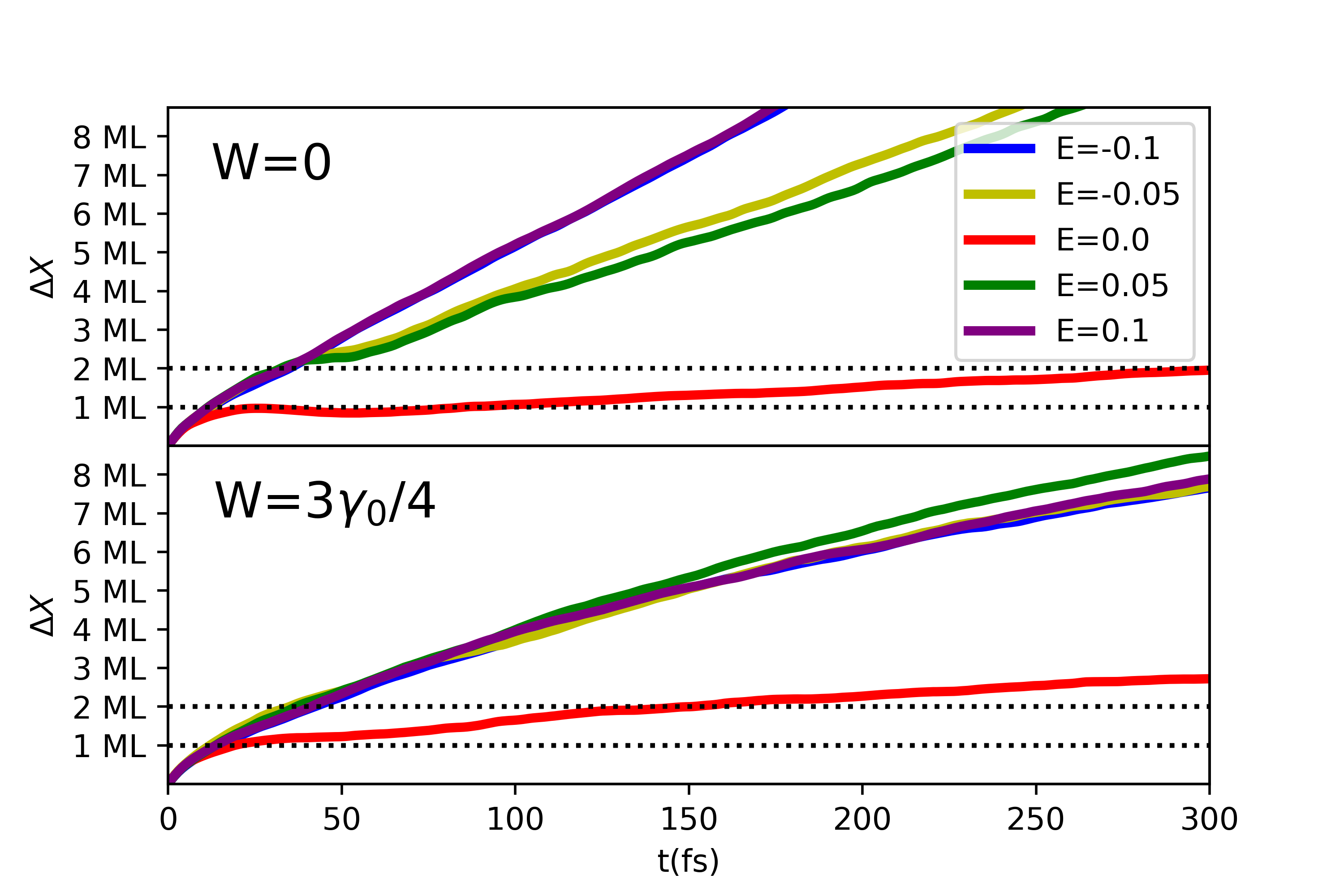}
\caption{Electronic spreading, in units of the moiré length (ML), as a function of time for the clean system (upper panel) and the disordered system with Anderson disorder $W=3\gamma_0/4$ (lower panel). Black dotted lines indicate spreadings corresponding to 1 and 2 ML.}
\label{figSupEWP1}
\end{figure}

In this section we illustrate how transport is impacted by the superperiodicity of the moiré lattice in the short-time limit. In Fig.\ \ref{figSupEWP1} we show the spreading of the wave packet at short times, given by $\Delta X = \sqrt{\Delta X^2}$, where $\Delta X^2$ is the mean square displacement as defined in Eq.\ (4) of the main text. This quantity has been previously related with the real space evolution of the wave function for anisotropic systems \cite{alcón2023tailoring}. The top panel shows the ballistic case and the bottom panel is with disorder strength $W = 3\gamma_0/4$. On the $y$-axis, the wave packet spreading is shown in units of the moiré length of the MATBLG system, which is $\sim$13 nm.

In the ballistic case (top panel), for very short times before reaching the first moiré length (ML), states propagate with the same velocity at all energies. However, upon reaching the first ML, the low-energy states in the flat band (red curve) undergo a dramatic slowing of their propagation velocity. States adjacent to the flat band (yellow and green curves) experience a slowing at a spreading of 2ML, while high-energy states (blue and purple curves) experience little velocity renormalization.

In the disordered case (bottom panel), similar behavior is seen for the flat band (red curve), while states at all other energies converge to the same behavior owing to the Anderson disorder.

These results present a direct visualization of the accommodation of the wave function to the moiré lattice and the complex flat-band physics of MATBLG. They also provide lower spatial and temporal bounds for the appearance of such flat-band physics, which do not appear until the wave function has evolved long enough to see the superiodicity of the moiré lattice.

%



\begin{thebibliography}{79}%
\makeatletter
\providecommand \@ifxundefined [1]{%
 \@ifx{#1\undefined}
}%
\providecommand \@ifnum [1]{%
 \ifnum #1\expandafter \@firstoftwo
 \else \expandafter \@secondoftwo
 \fi
}%
\providecommand \@ifx [1]{%
 \ifx #1\expandafter \@firstoftwo
 \else \expandafter \@secondoftwo
 \fi
}%
\providecommand \natexlab [1]{#1}%
\providecommand \enquote  [1]{``#1''}%
\providecommand \bibnamefont  [1]{#1}%
\providecommand \bibfnamefont [1]{#1}%
\providecommand \citenamefont [1]{#1}%
\providecommand \href@noop [0]{\@secondoftwo}%
\providecommand \href [0]{\begingroup \@sanitize@url \@href}%
\providecommand \@href[1]{\@@startlink{#1}\@@href}%
\providecommand \@@href[1]{\endgroup#1\@@endlink}%
\providecommand \@sanitize@url [0]{\catcode `\\12\catcode `\$12\catcode `\&12\catcode `\#12\catcode `\^12\catcode `\_12\catcode `\%12\relax}%
\providecommand \@@startlink[1]{}%
\providecommand \@@endlink[0]{}%
\providecommand \url  [0]{\begingroup\@sanitize@url \@url }%
\providecommand \@url [1]{\endgroup\@href {#1}{\urlprefix }}%
\providecommand \urlprefix  [0]{URL }%
\providecommand \Eprint [0]{\href }%
\providecommand \doibase [0]{https://doi.org/}%
\providecommand \selectlanguage [0]{\@gobble}%
\providecommand \bibinfo  [0]{\@secondoftwo}%
\providecommand \bibfield  [0]{\@secondoftwo}%
\providecommand \translation [1]{[#1]}%
\providecommand \BibitemOpen [0]{}%
\providecommand \bibitemStop [0]{}%
\providecommand \bibitemNoStop [0]{.\EOS\space}%
\providecommand \EOS [0]{\spacefactor3000\relax}%
\providecommand \BibitemShut  [1]{\csname bibitem#1\endcsname}%
\let\auto@bib@innerbib\@empty
\bibitem [{\citenamefont {Cao}\ \emph {et~al.}(2018{\natexlab{a}})\citenamefont {Cao}, \citenamefont {Fatemi}, \citenamefont {Fang}, \citenamefont {Watanabe}, \citenamefont {Taniguchi}, \citenamefont {Kaxiras},\ and\ \citenamefont {Jarillo-Herrero}}]{Cao2018}%
  \BibitemOpen
  \bibfield  {author} {\bibinfo {author} {\bibfnamefont {Y.}~\bibnamefont {Cao}}, \bibinfo {author} {\bibfnamefont {V.}~\bibnamefont {Fatemi}}, \bibinfo {author} {\bibfnamefont {S.}~\bibnamefont {Fang}}, \bibinfo {author} {\bibfnamefont {K.}~\bibnamefont {Watanabe}}, \bibinfo {author} {\bibfnamefont {T.}~\bibnamefont {Taniguchi}}, \bibinfo {author} {\bibfnamefont {E.}~\bibnamefont {Kaxiras}},\ and\ \bibinfo {author} {\bibfnamefont {P.}~\bibnamefont {Jarillo-Herrero}},\ }\bibfield  {title} {\bibinfo {title} {Unconventional superconductivity in magic-angle graphene superlattices},\ }\href {https://doi.org/10.1038/nature26160} {\bibfield  {journal} {\bibinfo  {journal} {Nature}\ }\textbf {\bibinfo {volume} {556}},\ \bibinfo {pages} {43} (\bibinfo {year} {2018}{\natexlab{a}})}\BibitemShut {NoStop}%
\bibitem [{\citenamefont {Cao}\ \emph {et~al.}(2018{\natexlab{b}})\citenamefont {Cao}, \citenamefont {Fatemi}, \citenamefont {Demir}, \citenamefont {Fang}, \citenamefont {Tomarken}, \citenamefont {Luo}, \citenamefont {Sanchez-Yamagishi}, \citenamefont {Watanabe}, \citenamefont {Taniguchi}, \citenamefont {Kaxiras}, \citenamefont {Ashoori},\ and\ \citenamefont {Jarillo-Herrero}}]{Cao2018b}%
  \BibitemOpen
  \bibfield  {author} {\bibinfo {author} {\bibfnamefont {Y.}~\bibnamefont {Cao}}, \bibinfo {author} {\bibfnamefont {V.}~\bibnamefont {Fatemi}}, \bibinfo {author} {\bibfnamefont {A.}~\bibnamefont {Demir}}, \bibinfo {author} {\bibfnamefont {S.}~\bibnamefont {Fang}}, \bibinfo {author} {\bibfnamefont {S.~L.}\ \bibnamefont {Tomarken}}, \bibinfo {author} {\bibfnamefont {J.~Y.}\ \bibnamefont {Luo}}, \bibinfo {author} {\bibfnamefont {J.~D.}\ \bibnamefont {Sanchez-Yamagishi}}, \bibinfo {author} {\bibfnamefont {K.}~\bibnamefont {Watanabe}}, \bibinfo {author} {\bibfnamefont {T.}~\bibnamefont {Taniguchi}}, \bibinfo {author} {\bibfnamefont {E.}~\bibnamefont {Kaxiras}}, \bibinfo {author} {\bibfnamefont {R.~C.}\ \bibnamefont {Ashoori}},\ and\ \bibinfo {author} {\bibfnamefont {P.}~\bibnamefont {Jarillo-Herrero}},\ }\bibfield  {title} {\bibinfo {title} {Correlated insulator behaviour at half-filling in magic-angle graphene superlattices},\ }\href {https://doi.org/10.1038/nature26154} {\bibfield  {journal} {\bibinfo  {journal}
  {Nature}\ }\textbf {\bibinfo {volume} {556}},\ \bibinfo {pages} {80} (\bibinfo {year} {2018}{\natexlab{b}})}\BibitemShut {NoStop}%
\bibitem [{\citenamefont {Stepanov}\ \emph {et~al.}(2020)\citenamefont {Stepanov}, \citenamefont {Das}, \citenamefont {Lu}, \citenamefont {Fahimniya}, \citenamefont {Watanabe}, \citenamefont {Taniguchi}, \citenamefont {Koppens}, \citenamefont {Lischner}, \citenamefont {Levitov},\ and\ \citenamefont {Efetov}}]{Stepanov2020}%
  \BibitemOpen
  \bibfield  {author} {\bibinfo {author} {\bibfnamefont {P.}~\bibnamefont {Stepanov}}, \bibinfo {author} {\bibfnamefont {I.}~\bibnamefont {Das}}, \bibinfo {author} {\bibfnamefont {X.}~\bibnamefont {Lu}}, \bibinfo {author} {\bibfnamefont {A.}~\bibnamefont {Fahimniya}}, \bibinfo {author} {\bibfnamefont {K.}~\bibnamefont {Watanabe}}, \bibinfo {author} {\bibfnamefont {T.}~\bibnamefont {Taniguchi}}, \bibinfo {author} {\bibfnamefont {F.~H.~L.}\ \bibnamefont {Koppens}}, \bibinfo {author} {\bibfnamefont {J.}~\bibnamefont {Lischner}}, \bibinfo {author} {\bibfnamefont {L.}~\bibnamefont {Levitov}},\ and\ \bibinfo {author} {\bibfnamefont {D.~K.}\ \bibnamefont {Efetov}},\ }\bibfield  {title} {\bibinfo {title} {Untying the insulating and superconducting orders in magic-angle graphene},\ }\href {https://doi.org/10.1038/s41586-020-2459-6} {\bibfield  {journal} {\bibinfo  {journal} {Nature}\ }\textbf {\bibinfo {volume} {583}},\ \bibinfo {pages} {375} (\bibinfo {year} {2020})}\BibitemShut {NoStop}%
\bibitem [{\citenamefont {Bistritzer}\ and\ \citenamefont {MacDonald}(2011)}]{pnas1108174108}%
  \BibitemOpen
  \bibfield  {author} {\bibinfo {author} {\bibfnamefont {R.}~\bibnamefont {Bistritzer}}\ and\ \bibinfo {author} {\bibfnamefont {A.~H.}\ \bibnamefont {MacDonald}},\ }\bibfield  {title} {\bibinfo {title} {Moiré bands in twisted double-layer graphene},\ }\href {https://doi.org/10.1073/pnas.1108174108} {\bibfield  {journal} {\bibinfo  {journal} {Proc. Natl. Acad. Sci. U.S.A.}\ }\textbf {\bibinfo {volume} {108}},\ \bibinfo {pages} {12233} (\bibinfo {year} {2011})}\BibitemShut {NoStop}%
\bibitem [{\citenamefont {Andrei}\ \emph {et~al.}(2021)\citenamefont {Andrei}, \citenamefont {Efetov}, \citenamefont {Jarillo-Herrero}, \citenamefont {MacDonald}, \citenamefont {Mak}, \citenamefont {Senthil}, \citenamefont {Tutuc}, \citenamefont {Yazdani},\ and\ \citenamefont {Young}}]{Andrei2021}%
  \BibitemOpen
  \bibfield  {author} {\bibinfo {author} {\bibfnamefont {E.~Y.}\ \bibnamefont {Andrei}}, \bibinfo {author} {\bibfnamefont {D.~K.}\ \bibnamefont {Efetov}}, \bibinfo {author} {\bibfnamefont {P.}~\bibnamefont {Jarillo-Herrero}}, \bibinfo {author} {\bibfnamefont {A.~H.}\ \bibnamefont {MacDonald}}, \bibinfo {author} {\bibfnamefont {K.~F.}\ \bibnamefont {Mak}}, \bibinfo {author} {\bibfnamefont {T.}~\bibnamefont {Senthil}}, \bibinfo {author} {\bibfnamefont {E.}~\bibnamefont {Tutuc}}, \bibinfo {author} {\bibfnamefont {A.}~\bibnamefont {Yazdani}},\ and\ \bibinfo {author} {\bibfnamefont {A.~F.}\ \bibnamefont {Young}},\ }\bibfield  {title} {\bibinfo {title} {The marvels of moir{\'e} materials},\ }\href {https://doi.org/10.1038/s41578-021-00284-1} {\bibfield  {journal} {\bibinfo  {journal} {Nat. Rev. Mater.}\ }\textbf {\bibinfo {volume} {6}},\ \bibinfo {pages} {201} (\bibinfo {year} {2021})}\BibitemShut {NoStop}%
\bibitem [{\citenamefont {Tilak}\ \emph {et~al.}(2021)\citenamefont {Tilak}, \citenamefont {Lai}, \citenamefont {Wu}, \citenamefont {Zhang}, \citenamefont {Xu}, \citenamefont {Ribeiro}, \citenamefont {Canfield},\ and\ \citenamefont {Andrei}}]{Tilak2021}%
  \BibitemOpen
  \bibfield  {author} {\bibinfo {author} {\bibfnamefont {N.}~\bibnamefont {Tilak}}, \bibinfo {author} {\bibfnamefont {X.}~\bibnamefont {Lai}}, \bibinfo {author} {\bibfnamefont {S.}~\bibnamefont {Wu}}, \bibinfo {author} {\bibfnamefont {Z.}~\bibnamefont {Zhang}}, \bibinfo {author} {\bibfnamefont {M.}~\bibnamefont {Xu}}, \bibinfo {author} {\bibfnamefont {R.~d.~A.}\ \bibnamefont {Ribeiro}}, \bibinfo {author} {\bibfnamefont {P.~C.}\ \bibnamefont {Canfield}},\ and\ \bibinfo {author} {\bibfnamefont {E.~Y.}\ \bibnamefont {Andrei}},\ }\bibfield  {title} {\bibinfo {title} {Flat band carrier confinement in magic-angle twisted bilayer graphene},\ }\href {https://doi.org/10.1038/s41467-021-24480-3} {\bibfield  {journal} {\bibinfo  {journal} {Nat. Commun.}\ }\textbf {\bibinfo {volume} {12}},\ \bibinfo {pages} {4180} (\bibinfo {year} {2021})}\BibitemShut {NoStop}%
\bibitem [{\citenamefont {Serlin}\ \emph {et~al.}(2020)\citenamefont {Serlin}, \citenamefont {Tschirhart}, \citenamefont {Polshyn}, \citenamefont {Zhang}, \citenamefont {Zhu}, \citenamefont {Watanabe}, \citenamefont {Taniguchi}, \citenamefont {Balents},\ and\ \citenamefont {Young}}]{Serlin2020}%
  \BibitemOpen
  \bibfield  {author} {\bibinfo {author} {\bibfnamefont {M.}~\bibnamefont {Serlin}}, \bibinfo {author} {\bibfnamefont {C.~L.}\ \bibnamefont {Tschirhart}}, \bibinfo {author} {\bibfnamefont {H.}~\bibnamefont {Polshyn}}, \bibinfo {author} {\bibfnamefont {Y.}~\bibnamefont {Zhang}}, \bibinfo {author} {\bibfnamefont {J.}~\bibnamefont {Zhu}}, \bibinfo {author} {\bibfnamefont {K.}~\bibnamefont {Watanabe}}, \bibinfo {author} {\bibfnamefont {T.}~\bibnamefont {Taniguchi}}, \bibinfo {author} {\bibfnamefont {L.}~\bibnamefont {Balents}},\ and\ \bibinfo {author} {\bibfnamefont {A.~F.}\ \bibnamefont {Young}},\ }\bibfield  {title} {\bibinfo {title} {Intrinsic quantized anomalous hall effect in a moiré heterostructure},\ }\href {https://doi.org/10.1126/science.aay5533} {\bibfield  {journal} {\bibinfo  {journal} {Science}\ }\textbf {\bibinfo {volume} {367}},\ \bibinfo {pages} {900} (\bibinfo {year} {2020})}\BibitemShut {NoStop}%
\bibitem [{\citenamefont {Saito}\ \emph {et~al.}(2020)\citenamefont {Saito}, \citenamefont {Ge}, \citenamefont {Watanabe}, \citenamefont {Taniguchi},\ and\ \citenamefont {Young}}]{Saito2020}%
  \BibitemOpen
  \bibfield  {author} {\bibinfo {author} {\bibfnamefont {Y.}~\bibnamefont {Saito}}, \bibinfo {author} {\bibfnamefont {J.}~\bibnamefont {Ge}}, \bibinfo {author} {\bibfnamefont {K.}~\bibnamefont {Watanabe}}, \bibinfo {author} {\bibfnamefont {T.}~\bibnamefont {Taniguchi}},\ and\ \bibinfo {author} {\bibfnamefont {A.~F.}\ \bibnamefont {Young}},\ }\bibfield  {title} {\bibinfo {title} {Independent superconductors and correlated insulators in twisted bilayer graphene},\ }\href {https://doi.org/10.1038/s41567-020-0928-3} {\bibfield  {journal} {\bibinfo  {journal} {Nat. Phys.}\ }\textbf {\bibinfo {volume} {16}},\ \bibinfo {pages} {926} (\bibinfo {year} {2020})}\BibitemShut {NoStop}%
\bibitem [{\citenamefont {Balents}\ \emph {et~al.}(2020)\citenamefont {Balents}, \citenamefont {Dean}, \citenamefont {Efetov},\ and\ \citenamefont {Young}}]{Balents2020}%
  \BibitemOpen
  \bibfield  {author} {\bibinfo {author} {\bibfnamefont {L.}~\bibnamefont {Balents}}, \bibinfo {author} {\bibfnamefont {C.~R.}\ \bibnamefont {Dean}}, \bibinfo {author} {\bibfnamefont {D.~K.}\ \bibnamefont {Efetov}},\ and\ \bibinfo {author} {\bibfnamefont {A.~F.}\ \bibnamefont {Young}},\ }\bibfield  {title} {\bibinfo {title} {Superconductivity and strong correlations in moir{\'e} flat bands},\ }\href {https://doi.org/10.1038/s41567-020-0906-9} {\bibfield  {journal} {\bibinfo  {journal} {Nat. Phys.}\ }\textbf {\bibinfo {volume} {16}},\ \bibinfo {pages} {725} (\bibinfo {year} {2020})}\BibitemShut {NoStop}%
\bibitem [{\citenamefont {Stepanov}\ \emph {et~al.}(2021)\citenamefont {Stepanov}, \citenamefont {Xie}, \citenamefont {Taniguchi}, \citenamefont {Watanabe}, \citenamefont {Lu}, \citenamefont {MacDonald}, \citenamefont {Bernevig},\ and\ \citenamefont {Efetov}}]{PhysRevLett.127.197701}%
  \BibitemOpen
  \bibfield  {author} {\bibinfo {author} {\bibfnamefont {P.}~\bibnamefont {Stepanov}}, \bibinfo {author} {\bibfnamefont {M.}~\bibnamefont {Xie}}, \bibinfo {author} {\bibfnamefont {T.}~\bibnamefont {Taniguchi}}, \bibinfo {author} {\bibfnamefont {K.}~\bibnamefont {Watanabe}}, \bibinfo {author} {\bibfnamefont {X.}~\bibnamefont {Lu}}, \bibinfo {author} {\bibfnamefont {A.~H.}\ \bibnamefont {MacDonald}}, \bibinfo {author} {\bibfnamefont {B.~A.}\ \bibnamefont {Bernevig}},\ and\ \bibinfo {author} {\bibfnamefont {D.~K.}\ \bibnamefont {Efetov}},\ }\bibfield  {title} {\bibinfo {title} {Competing zero-field chern insulators in superconducting twisted bilayer graphene},\ }\href {https://doi.org/10.1103/PhysRevLett.127.197701} {\bibfield  {journal} {\bibinfo  {journal} {Phys. Rev. Lett.}\ }\textbf {\bibinfo {volume} {127}},\ \bibinfo {pages} {197701} (\bibinfo {year} {2021})}\BibitemShut {NoStop}%
\bibitem [{\citenamefont {Xie}\ \emph {et~al.}(2021)\citenamefont {Xie}, \citenamefont {Pierce}, \citenamefont {Park}, \citenamefont {Parker}, \citenamefont {Khalaf}, \citenamefont {Ledwith}, \citenamefont {Cao}, \citenamefont {Lee}, \citenamefont {Chen}, \citenamefont {Forrester}, \citenamefont {Watanabe}, \citenamefont {Taniguchi}, \citenamefont {Vishwanath}, \citenamefont {Jarillo-Herrero},\ and\ \citenamefont {Yacoby}}]{Xie2021}%
  \BibitemOpen
  \bibfield  {author} {\bibinfo {author} {\bibfnamefont {Y.}~\bibnamefont {Xie}}, \bibinfo {author} {\bibfnamefont {A.~T.}\ \bibnamefont {Pierce}}, \bibinfo {author} {\bibfnamefont {J.~M.}\ \bibnamefont {Park}}, \bibinfo {author} {\bibfnamefont {D.~E.}\ \bibnamefont {Parker}}, \bibinfo {author} {\bibfnamefont {E.}~\bibnamefont {Khalaf}}, \bibinfo {author} {\bibfnamefont {P.}~\bibnamefont {Ledwith}}, \bibinfo {author} {\bibfnamefont {Y.}~\bibnamefont {Cao}}, \bibinfo {author} {\bibfnamefont {S.~H.}\ \bibnamefont {Lee}}, \bibinfo {author} {\bibfnamefont {S.}~\bibnamefont {Chen}}, \bibinfo {author} {\bibfnamefont {P.~R.}\ \bibnamefont {Forrester}}, \bibinfo {author} {\bibfnamefont {K.}~\bibnamefont {Watanabe}}, \bibinfo {author} {\bibfnamefont {T.}~\bibnamefont {Taniguchi}}, \bibinfo {author} {\bibfnamefont {A.}~\bibnamefont {Vishwanath}}, \bibinfo {author} {\bibfnamefont {P.}~\bibnamefont {Jarillo-Herrero}},\ and\ \bibinfo {author} {\bibfnamefont {A.}~\bibnamefont {Yacoby}},\ }\bibfield  {title} {\bibinfo
  {title} {Fractional chern insulators in magic-angle twisted bilayer graphene},\ }\href {https://doi.org/10.1038/s41586-021-04002-3} {\bibfield  {journal} {\bibinfo  {journal} {Nature}\ }\textbf {\bibinfo {volume} {600}},\ \bibinfo {pages} {439} (\bibinfo {year} {2021})}\BibitemShut {NoStop}%
\bibitem [{\citenamefont {Pierce}\ \emph {et~al.}(2021)\citenamefont {Pierce}, \citenamefont {Xie}, \citenamefont {Park}, \citenamefont {Khalaf}, \citenamefont {Lee}, \citenamefont {Cao}, \citenamefont {Parker}, \citenamefont {Forrester}, \citenamefont {Chen}, \citenamefont {Watanabe}, \citenamefont {Taniguchi}, \citenamefont {Vishwanath}, \citenamefont {Jarillo-Herrero},\ and\ \citenamefont {Yacoby}}]{Pierce2021}%
  \BibitemOpen
  \bibfield  {author} {\bibinfo {author} {\bibfnamefont {A.~T.}\ \bibnamefont {Pierce}}, \bibinfo {author} {\bibfnamefont {Y.}~\bibnamefont {Xie}}, \bibinfo {author} {\bibfnamefont {J.~M.}\ \bibnamefont {Park}}, \bibinfo {author} {\bibfnamefont {E.}~\bibnamefont {Khalaf}}, \bibinfo {author} {\bibfnamefont {S.~H.}\ \bibnamefont {Lee}}, \bibinfo {author} {\bibfnamefont {Y.}~\bibnamefont {Cao}}, \bibinfo {author} {\bibfnamefont {D.~E.}\ \bibnamefont {Parker}}, \bibinfo {author} {\bibfnamefont {P.~R.}\ \bibnamefont {Forrester}}, \bibinfo {author} {\bibfnamefont {S.}~\bibnamefont {Chen}}, \bibinfo {author} {\bibfnamefont {K.}~\bibnamefont {Watanabe}}, \bibinfo {author} {\bibfnamefont {T.}~\bibnamefont {Taniguchi}}, \bibinfo {author} {\bibfnamefont {A.}~\bibnamefont {Vishwanath}}, \bibinfo {author} {\bibfnamefont {P.}~\bibnamefont {Jarillo-Herrero}},\ and\ \bibinfo {author} {\bibfnamefont {A.}~\bibnamefont {Yacoby}},\ }\bibfield  {title} {\bibinfo {title} {Unconventional sequence of correlated chern insulators in
  magic-angle twisted bilayer graphene},\ }\href {https://doi.org/10.1038/s41567-021-01347-4} {\bibfield  {journal} {\bibinfo  {journal} {Nat. Phys.}\ }\textbf {\bibinfo {volume} {17}},\ \bibinfo {pages} {1210} (\bibinfo {year} {2021})}\BibitemShut {NoStop}%
\bibitem [{\citenamefont {Kwan}\ \emph {et~al.}(2021)\citenamefont {Kwan}, \citenamefont {Hu}, \citenamefont {Simon},\ and\ \citenamefont {Parameswaran}}]{PhysRevLett.126.137601}%
  \BibitemOpen
  \bibfield  {author} {\bibinfo {author} {\bibfnamefont {Y.~H.}\ \bibnamefont {Kwan}}, \bibinfo {author} {\bibfnamefont {Y.}~\bibnamefont {Hu}}, \bibinfo {author} {\bibfnamefont {S.~H.}\ \bibnamefont {Simon}},\ and\ \bibinfo {author} {\bibfnamefont {S.~A.}\ \bibnamefont {Parameswaran}},\ }\bibfield  {title} {\bibinfo {title} {Exciton band topology in spontaneous quantum anomalous hall insulators: Applications to twisted bilayer graphene},\ }\href {https://doi.org/10.1103/PhysRevLett.126.137601} {\bibfield  {journal} {\bibinfo  {journal} {Phys. Rev. Lett.}\ }\textbf {\bibinfo {volume} {126}},\ \bibinfo {pages} {137601} (\bibinfo {year} {2021})}\BibitemShut {NoStop}%
\bibitem [{\citenamefont {Park}\ \emph {et~al.}(2022)\citenamefont {Park}, \citenamefont {Cao}, \citenamefont {Xia}, \citenamefont {Sun}, \citenamefont {Watanabe}, \citenamefont {Taniguchi},\ and\ \citenamefont {Jarillo-Herrero}}]{Park2022}%
  \BibitemOpen
  \bibfield  {author} {\bibinfo {author} {\bibfnamefont {J.~M.}\ \bibnamefont {Park}}, \bibinfo {author} {\bibfnamefont {Y.}~\bibnamefont {Cao}}, \bibinfo {author} {\bibfnamefont {L.-Q.}\ \bibnamefont {Xia}}, \bibinfo {author} {\bibfnamefont {S.}~\bibnamefont {Sun}}, \bibinfo {author} {\bibfnamefont {K.}~\bibnamefont {Watanabe}}, \bibinfo {author} {\bibfnamefont {T.}~\bibnamefont {Taniguchi}},\ and\ \bibinfo {author} {\bibfnamefont {P.}~\bibnamefont {Jarillo-Herrero}},\ }\bibfield  {title} {\bibinfo {title} {Robust superconductivity in magic-angle multilayer graphene family},\ }\href {https://doi.org/10.1038/s41563-022-01287-1} {\bibfield  {journal} {\bibinfo  {journal} {Nat. Mater.}\ }\textbf {\bibinfo {volume} {21}},\ \bibinfo {pages} {877} (\bibinfo {year} {2022})}\BibitemShut {NoStop}%
\bibitem [{\citenamefont {Jaoui}\ \emph {et~al.}(2022)\citenamefont {Jaoui}, \citenamefont {Das}, \citenamefont {Di~Battista}, \citenamefont {D{\'i}ez-M{\'e}rida}, \citenamefont {Lu}, \citenamefont {Watanabe}, \citenamefont {Taniguchi}, \citenamefont {Ishizuka}, \citenamefont {Levitov},\ and\ \citenamefont {Efetov}}]{Jaoui2022}%
  \BibitemOpen
  \bibfield  {author} {\bibinfo {author} {\bibfnamefont {A.}~\bibnamefont {Jaoui}}, \bibinfo {author} {\bibfnamefont {I.}~\bibnamefont {Das}}, \bibinfo {author} {\bibfnamefont {G.}~\bibnamefont {Di~Battista}}, \bibinfo {author} {\bibfnamefont {J.}~\bibnamefont {D{\'i}ez-M{\'e}rida}}, \bibinfo {author} {\bibfnamefont {X.}~\bibnamefont {Lu}}, \bibinfo {author} {\bibfnamefont {K.}~\bibnamefont {Watanabe}}, \bibinfo {author} {\bibfnamefont {T.}~\bibnamefont {Taniguchi}}, \bibinfo {author} {\bibfnamefont {H.}~\bibnamefont {Ishizuka}}, \bibinfo {author} {\bibfnamefont {L.}~\bibnamefont {Levitov}},\ and\ \bibinfo {author} {\bibfnamefont {D.~K.}\ \bibnamefont {Efetov}},\ }\bibfield  {title} {\bibinfo {title} {Quantum critical behaviour in magic-angle twisted bilayer graphene},\ }\href {https://doi.org/10.1038/s41567-022-01556-5} {\bibfield  {journal} {\bibinfo  {journal} {Nat. Phys.}\ }\textbf {\bibinfo {volume} {18}},\ \bibinfo {pages} {633} (\bibinfo {year} {2022})}\BibitemShut {NoStop}%
\bibitem [{\citenamefont {Paul}\ \emph {et~al.}(2022)\citenamefont {Paul}, \citenamefont {Ghosh}, \citenamefont {Chakraborty}, \citenamefont {Roy}, \citenamefont {Dutta}, \citenamefont {Watanabe}, \citenamefont {Taniguchi}, \citenamefont {Panda}, \citenamefont {Agarwala}, \citenamefont {Mukerjee}, \citenamefont {Banerjee},\ and\ \citenamefont {Das}}]{Paul2022}%
  \BibitemOpen
  \bibfield  {author} {\bibinfo {author} {\bibfnamefont {A.~K.}\ \bibnamefont {Paul}}, \bibinfo {author} {\bibfnamefont {A.}~\bibnamefont {Ghosh}}, \bibinfo {author} {\bibfnamefont {S.}~\bibnamefont {Chakraborty}}, \bibinfo {author} {\bibfnamefont {U.}~\bibnamefont {Roy}}, \bibinfo {author} {\bibfnamefont {R.}~\bibnamefont {Dutta}}, \bibinfo {author} {\bibfnamefont {K.}~\bibnamefont {Watanabe}}, \bibinfo {author} {\bibfnamefont {T.}~\bibnamefont {Taniguchi}}, \bibinfo {author} {\bibfnamefont {A.}~\bibnamefont {Panda}}, \bibinfo {author} {\bibfnamefont {A.}~\bibnamefont {Agarwala}}, \bibinfo {author} {\bibfnamefont {S.}~\bibnamefont {Mukerjee}}, \bibinfo {author} {\bibfnamefont {S.}~\bibnamefont {Banerjee}},\ and\ \bibinfo {author} {\bibfnamefont {A.}~\bibnamefont {Das}},\ }\bibfield  {title} {\bibinfo {title} {Interaction-driven giant thermopower in magic-angle twisted bilayer graphene},\ }\href {https://doi.org/10.1038/s41567-022-01574-3} {\bibfield  {journal} {\bibinfo  {journal} {Nat. Phys.}\ }\textbf
  {\bibinfo {volume} {18}},\ \bibinfo {pages} {691} (\bibinfo {year} {2022})}\BibitemShut {NoStop}%
\bibitem [{\citenamefont {Klein}\ \emph {et~al.}(2023)\citenamefont {Klein}, \citenamefont {Xia}, \citenamefont {MacNeill}, \citenamefont {Watanabe}, \citenamefont {Taniguchi},\ and\ \citenamefont {Jarillo-Herrero}}]{Klein2023}%
  \BibitemOpen
  \bibfield  {author} {\bibinfo {author} {\bibfnamefont {D.~R.}\ \bibnamefont {Klein}}, \bibinfo {author} {\bibfnamefont {L.-Q.}\ \bibnamefont {Xia}}, \bibinfo {author} {\bibfnamefont {D.}~\bibnamefont {MacNeill}}, \bibinfo {author} {\bibfnamefont {K.}~\bibnamefont {Watanabe}}, \bibinfo {author} {\bibfnamefont {T.}~\bibnamefont {Taniguchi}},\ and\ \bibinfo {author} {\bibfnamefont {P.}~\bibnamefont {Jarillo-Herrero}},\ }\bibfield  {title} {\bibinfo {title} {Electrical switching of a bistable moir{\'e} superconductor},\ }\href {https://doi.org/10.1038/s41565-022-01314-x} {\bibfield  {journal} {\bibinfo  {journal} {Nat. Nanotechnol.}\ }\textbf {\bibinfo {volume} {18}},\ \bibinfo {pages} {331} (\bibinfo {year} {2023})}\BibitemShut {NoStop}%
\bibitem [{\citenamefont {Tian}\ \emph {et~al.}(2023)\citenamefont {Tian}, \citenamefont {Gao}, \citenamefont {Zhang}, \citenamefont {Che}, \citenamefont {Xu}, \citenamefont {Cheung}, \citenamefont {Watanabe}, \citenamefont {Taniguchi}, \citenamefont {Randeria}, \citenamefont {Zhang}, \citenamefont {Lau},\ and\ \citenamefont {Bockrath}}]{Tian2023}%
  \BibitemOpen
  \bibfield  {author} {\bibinfo {author} {\bibfnamefont {H.}~\bibnamefont {Tian}}, \bibinfo {author} {\bibfnamefont {X.}~\bibnamefont {Gao}}, \bibinfo {author} {\bibfnamefont {Y.}~\bibnamefont {Zhang}}, \bibinfo {author} {\bibfnamefont {S.}~\bibnamefont {Che}}, \bibinfo {author} {\bibfnamefont {T.}~\bibnamefont {Xu}}, \bibinfo {author} {\bibfnamefont {P.}~\bibnamefont {Cheung}}, \bibinfo {author} {\bibfnamefont {K.}~\bibnamefont {Watanabe}}, \bibinfo {author} {\bibfnamefont {T.}~\bibnamefont {Taniguchi}}, \bibinfo {author} {\bibfnamefont {M.}~\bibnamefont {Randeria}}, \bibinfo {author} {\bibfnamefont {F.}~\bibnamefont {Zhang}}, \bibinfo {author} {\bibfnamefont {C.~N.}\ \bibnamefont {Lau}},\ and\ \bibinfo {author} {\bibfnamefont {M.~W.}\ \bibnamefont {Bockrath}},\ }\bibfield  {title} {\bibinfo {title} {Evidence for dirac flat band superconductivity enabled by quantum geometry},\ }\href {https://doi.org/10.1038/s41586-022-05576-2} {\bibfield  {journal} {\bibinfo  {journal} {Nature}\ }\textbf {\bibinfo {volume}
  {614}},\ \bibinfo {pages} {440} (\bibinfo {year} {2023})}\BibitemShut {NoStop}%
\bibitem [{\citenamefont {Ahn}\ \emph {et~al.}(2018)\citenamefont {Ahn}, \citenamefont {Moon}, \citenamefont {Kim}, \citenamefont {Kim}, \citenamefont {Shin}, \citenamefont {Kim}, \citenamefont {Cha}, \citenamefont {Kahng}, \citenamefont {Kim}, \citenamefont {Koshino}, \citenamefont {Son}, \citenamefont {Yang},\ and\ \citenamefont {Ahn}}]{Ahn2018}%
  \BibitemOpen
  \bibfield  {author} {\bibinfo {author} {\bibfnamefont {S.~J.}\ \bibnamefont {Ahn}}, \bibinfo {author} {\bibfnamefont {P.}~\bibnamefont {Moon}}, \bibinfo {author} {\bibfnamefont {T.-H.}\ \bibnamefont {Kim}}, \bibinfo {author} {\bibfnamefont {H.-W.}\ \bibnamefont {Kim}}, \bibinfo {author} {\bibfnamefont {H.-C.}\ \bibnamefont {Shin}}, \bibinfo {author} {\bibfnamefont {E.~H.}\ \bibnamefont {Kim}}, \bibinfo {author} {\bibfnamefont {H.~W.}\ \bibnamefont {Cha}}, \bibinfo {author} {\bibfnamefont {S.-J.}\ \bibnamefont {Kahng}}, \bibinfo {author} {\bibfnamefont {P.}~\bibnamefont {Kim}}, \bibinfo {author} {\bibfnamefont {M.}~\bibnamefont {Koshino}}, \bibinfo {author} {\bibfnamefont {Y.-W.}\ \bibnamefont {Son}}, \bibinfo {author} {\bibfnamefont {C.-W.}\ \bibnamefont {Yang}},\ and\ \bibinfo {author} {\bibfnamefont {J.~R.}\ \bibnamefont {Ahn}},\ }\bibfield  {title} {\bibinfo {title} {Dirac electrons in a dodecagonal graphene quasicrystal},\ }\href {https://doi.org/10.1126/science.aar8412} {\bibfield  {journal} {\bibinfo
  {journal} {Science}\ }\textbf {\bibinfo {volume} {361}},\ \bibinfo {pages} {782} (\bibinfo {year} {2018})}\BibitemShut {NoStop}%
\bibitem [{\citenamefont {Uri}\ \emph {et~al.}(2023)\citenamefont {Uri}, \citenamefont {de~la Barrera}, \citenamefont {Randeria}, \citenamefont {Rodan-Legrain}, \citenamefont {Devakul}, \citenamefont {Crowley}, \citenamefont {Paul}, \citenamefont {Watanabe}, \citenamefont {Taniguchi}, \citenamefont {Lifshitz}, \citenamefont {Fu}, \citenamefont {Ashoori},\ and\ \citenamefont {Jarillo-Herrero}}]{Uri2023}%
  \BibitemOpen
  \bibfield  {author} {\bibinfo {author} {\bibfnamefont {A.}~\bibnamefont {Uri}}, \bibinfo {author} {\bibfnamefont {S.~C.}\ \bibnamefont {de~la Barrera}}, \bibinfo {author} {\bibfnamefont {M.~T.}\ \bibnamefont {Randeria}}, \bibinfo {author} {\bibfnamefont {D.}~\bibnamefont {Rodan-Legrain}}, \bibinfo {author} {\bibfnamefont {T.}~\bibnamefont {Devakul}}, \bibinfo {author} {\bibfnamefont {P.~J.~D.}\ \bibnamefont {Crowley}}, \bibinfo {author} {\bibfnamefont {N.}~\bibnamefont {Paul}}, \bibinfo {author} {\bibfnamefont {K.}~\bibnamefont {Watanabe}}, \bibinfo {author} {\bibfnamefont {T.}~\bibnamefont {Taniguchi}}, \bibinfo {author} {\bibfnamefont {R.}~\bibnamefont {Lifshitz}}, \bibinfo {author} {\bibfnamefont {L.}~\bibnamefont {Fu}}, \bibinfo {author} {\bibfnamefont {R.~C.}\ \bibnamefont {Ashoori}},\ and\ \bibinfo {author} {\bibfnamefont {P.}~\bibnamefont {Jarillo-Herrero}},\ }\bibfield  {title} {\bibinfo {title} {Superconductivity and strong interactions in a tunable moir{\'e} quasicrystal},\ }\href
  {https://doi.org/10.1038/s41586-023-06294-z} {\bibfield  {journal} {\bibinfo  {journal} {Nature}\ }\textbf {\bibinfo {volume} {620}},\ \bibinfo {pages} {762} (\bibinfo {year} {2023})}\BibitemShut {NoStop}%
\bibitem [{\citenamefont {Lai}\ \emph {et~al.}(2023)\citenamefont {Lai}, \citenamefont {Guerci}, \citenamefont {Li}, \citenamefont {Watanabe}, \citenamefont {Taniguchi}, \citenamefont {Wilson}, \citenamefont {Pixley},\ and\ \citenamefont {Andrei}}]{lai2023imaging}%
  \BibitemOpen
  \bibfield  {author} {\bibinfo {author} {\bibfnamefont {X.}~\bibnamefont {Lai}}, \bibinfo {author} {\bibfnamefont {D.}~\bibnamefont {Guerci}}, \bibinfo {author} {\bibfnamefont {G.}~\bibnamefont {Li}}, \bibinfo {author} {\bibfnamefont {K.}~\bibnamefont {Watanabe}}, \bibinfo {author} {\bibfnamefont {T.}~\bibnamefont {Taniguchi}}, \bibinfo {author} {\bibfnamefont {J.}~\bibnamefont {Wilson}}, \bibinfo {author} {\bibfnamefont {J.~H.}\ \bibnamefont {Pixley}},\ and\ \bibinfo {author} {\bibfnamefont {E.~Y.}\ \bibnamefont {Andrei}},\ }\href@noop {} {\bibinfo {title} {Imaging self-aligned moir\'e crystals and quasicrystals in magic-angle bilayer graphene on hbn heterostructures}} (\bibinfo {year} {2023}),\ \Eprint {https://arxiv.org/abs/2311.07819} {arXiv:2311.07819 [cond-mat.mes-hall]} \BibitemShut {NoStop}%
\bibitem [{\citenamefont {Ledwith}\ \emph {et~al.}(2020)\citenamefont {Ledwith}, \citenamefont {Tarnopolsky}, \citenamefont {Khalaf},\ and\ \citenamefont {Vishwanath}}]{PhysRevResearch.2.023237}%
  \BibitemOpen
  \bibfield  {author} {\bibinfo {author} {\bibfnamefont {P.~J.}\ \bibnamefont {Ledwith}}, \bibinfo {author} {\bibfnamefont {G.}~\bibnamefont {Tarnopolsky}}, \bibinfo {author} {\bibfnamefont {E.}~\bibnamefont {Khalaf}},\ and\ \bibinfo {author} {\bibfnamefont {A.}~\bibnamefont {Vishwanath}},\ }\bibfield  {title} {\bibinfo {title} {Fractional chern insulator states in twisted bilayer graphene: An analytical approach},\ }\href {https://doi.org/10.1103/PhysRevResearch.2.023237} {\bibfield  {journal} {\bibinfo  {journal} {Phys. Rev. Res.}\ }\textbf {\bibinfo {volume} {2}},\ \bibinfo {pages} {023237} (\bibinfo {year} {2020})}\BibitemShut {NoStop}%
\bibitem [{\citenamefont {Ledwith}\ \emph {et~al.}(2021)\citenamefont {Ledwith}, \citenamefont {Khalaf},\ and\ \citenamefont {Vishwanath}}]{LEDWITH2021168646}%
  \BibitemOpen
  \bibfield  {author} {\bibinfo {author} {\bibfnamefont {P.~J.}\ \bibnamefont {Ledwith}}, \bibinfo {author} {\bibfnamefont {E.}~\bibnamefont {Khalaf}},\ and\ \bibinfo {author} {\bibfnamefont {A.}~\bibnamefont {Vishwanath}},\ }\bibfield  {title} {\bibinfo {title} {Strong coupling theory of magic-angle graphene: A pedagogical introduction},\ }\href {https://doi.org/https://doi.org/10.1016/j.aop.2021.168646} {\bibfield  {journal} {\bibinfo  {journal} {Ann. Phys.}\ }\textbf {\bibinfo {volume} {435}},\ \bibinfo {pages} {168646} (\bibinfo {year} {2021})},\ \bibinfo {note} {special issue on Philip W. Anderson}\BibitemShut {NoStop}%
\bibitem [{\citenamefont {Khalaf}\ \emph {et~al.}(2022)\citenamefont {Khalaf}, \citenamefont {Ledwith},\ and\ \citenamefont {Vishwanath}}]{PhysRevB.105.224508}%
  \BibitemOpen
  \bibfield  {author} {\bibinfo {author} {\bibfnamefont {E.}~\bibnamefont {Khalaf}}, \bibinfo {author} {\bibfnamefont {P.}~\bibnamefont {Ledwith}},\ and\ \bibinfo {author} {\bibfnamefont {A.}~\bibnamefont {Vishwanath}},\ }\bibfield  {title} {\bibinfo {title} {Symmetry constraints on superconductivity in twisted bilayer graphene: Fractional vortices, $4e$ condensates, or nonunitary pairing},\ }\href {https://doi.org/10.1103/PhysRevB.105.224508} {\bibfield  {journal} {\bibinfo  {journal} {Phys. Rev. B}\ }\textbf {\bibinfo {volume} {105}},\ \bibinfo {pages} {224508} (\bibinfo {year} {2022})}\BibitemShut {NoStop}%
\bibitem [{\citenamefont {Ciepielewski}\ \emph {et~al.}(2024)\citenamefont {Ciepielewski}, \citenamefont {Tworzydło}, \citenamefont {Hyart},\ and\ \citenamefont {Lau}}]{ciepielewski2024transport}%
  \BibitemOpen
  \bibfield  {author} {\bibinfo {author} {\bibfnamefont {A.~S.}\ \bibnamefont {Ciepielewski}}, \bibinfo {author} {\bibfnamefont {J.}~\bibnamefont {Tworzydło}}, \bibinfo {author} {\bibfnamefont {T.}~\bibnamefont {Hyart}},\ and\ \bibinfo {author} {\bibfnamefont {A.}~\bibnamefont {Lau}},\ }\href@noop {} {\bibinfo {title} {Transport effects of twist-angle disorder in mesoscopic twisted bilayer graphene}} (\bibinfo {year} {2024}),\ \Eprint {https://arxiv.org/abs/2403.19313} {arXiv:2403.19313 [cond-mat.mes-hall]} \BibitemShut {NoStop}%
\bibitem [{\citenamefont {Guinea}\ and\ \citenamefont {Walet}(2019)}]{PhysRevB.99.205134}%
  \BibitemOpen
  \bibfield  {author} {\bibinfo {author} {\bibfnamefont {F.}~\bibnamefont {Guinea}}\ and\ \bibinfo {author} {\bibfnamefont {N.~R.}\ \bibnamefont {Walet}},\ }\bibfield  {title} {\bibinfo {title} {Continuum models for twisted bilayer graphene: Effect of lattice deformation and hopping parameters},\ }\href {https://doi.org/10.1103/PhysRevB.99.205134} {\bibfield  {journal} {\bibinfo  {journal} {Phys. Rev. B}\ }\textbf {\bibinfo {volume} {99}},\ \bibinfo {pages} {205134} (\bibinfo {year} {2019})}\BibitemShut {NoStop}%
\bibitem [{\citenamefont {Nguyen}\ \emph {et~al.}(2021)\citenamefont {Nguyen}, \citenamefont {Paszko}, \citenamefont {Lamparski}, \citenamefont {Troeye}, \citenamefont {Meunier},\ and\ \citenamefont {Charlier}}]{Hung2021}%
  \BibitemOpen
  \bibfield  {author} {\bibinfo {author} {\bibfnamefont {V.~H.}\ \bibnamefont {Nguyen}}, \bibinfo {author} {\bibfnamefont {D.}~\bibnamefont {Paszko}}, \bibinfo {author} {\bibfnamefont {M.}~\bibnamefont {Lamparski}}, \bibinfo {author} {\bibfnamefont {B.~V.}\ \bibnamefont {Troeye}}, \bibinfo {author} {\bibfnamefont {V.}~\bibnamefont {Meunier}},\ and\ \bibinfo {author} {\bibfnamefont {J.-C.}\ \bibnamefont {Charlier}},\ }\bibfield  {title} {\bibinfo {title} {Electronic localization in small-angle twisted bilayer graphene},\ }\href {https://doi.org/10.1088/2053-1583/ac044f} {\bibfield  {journal} {\bibinfo  {journal} {2D Mater.}\ }\textbf {\bibinfo {volume} {8}},\ \bibinfo {pages} {035046} (\bibinfo {year} {2021})}\BibitemShut {NoStop}%
\bibitem [{\citenamefont {Mesple}\ \emph {et~al.}(2021)\citenamefont {Mesple}, \citenamefont {Missaoui}, \citenamefont {Cea}, \citenamefont {Huder}, \citenamefont {Guinea}, \citenamefont {Trambly~de Laissardi\`ere}, \citenamefont {Chapelier},\ and\ \citenamefont {Renard}}]{PhysRevLett.127.126405}%
  \BibitemOpen
  \bibfield  {author} {\bibinfo {author} {\bibfnamefont {F.}~\bibnamefont {Mesple}}, \bibinfo {author} {\bibfnamefont {A.}~\bibnamefont {Missaoui}}, \bibinfo {author} {\bibfnamefont {T.}~\bibnamefont {Cea}}, \bibinfo {author} {\bibfnamefont {L.}~\bibnamefont {Huder}}, \bibinfo {author} {\bibfnamefont {F.}~\bibnamefont {Guinea}}, \bibinfo {author} {\bibfnamefont {G.}~\bibnamefont {Trambly~de Laissardi\`ere}}, \bibinfo {author} {\bibfnamefont {C.}~\bibnamefont {Chapelier}},\ and\ \bibinfo {author} {\bibfnamefont {V.~T.}\ \bibnamefont {Renard}},\ }\bibfield  {title} {\bibinfo {title} {Heterostrain determines flat bands in magic-angle twisted graphene layers},\ }\href {https://doi.org/10.1103/PhysRevLett.127.126405} {\bibfield  {journal} {\bibinfo  {journal} {Phys. Rev. Lett.}\ }\textbf {\bibinfo {volume} {127}},\ \bibinfo {pages} {126405} (\bibinfo {year} {2021})}\BibitemShut {NoStop}%
\bibitem [{\citenamefont {Li}\ \emph {et~al.}(2022)\citenamefont {Li}, \citenamefont {Dong}, \citenamefont {Longhi}, \citenamefont {Liang}, \citenamefont {Xie},\ and\ \citenamefont {Yan}}]{PhysRevLett.129.220403}%
  \BibitemOpen
  \bibfield  {author} {\bibinfo {author} {\bibfnamefont {H.}~\bibnamefont {Li}}, \bibinfo {author} {\bibfnamefont {Z.}~\bibnamefont {Dong}}, \bibinfo {author} {\bibfnamefont {S.}~\bibnamefont {Longhi}}, \bibinfo {author} {\bibfnamefont {Q.}~\bibnamefont {Liang}}, \bibinfo {author} {\bibfnamefont {D.}~\bibnamefont {Xie}},\ and\ \bibinfo {author} {\bibfnamefont {B.}~\bibnamefont {Yan}},\ }\bibfield  {title} {\bibinfo {title} {Aharonov-bohm caging and inverse anderson transition in ultracold atoms},\ }\href {https://doi.org/10.1103/PhysRevLett.129.220403} {\bibfield  {journal} {\bibinfo  {journal} {Phys. Rev. Lett.}\ }\textbf {\bibinfo {volume} {129}},\ \bibinfo {pages} {220403} (\bibinfo {year} {2022})}\BibitemShut {NoStop}%
\bibitem [{\citenamefont {Bouzerar}\ and\ \citenamefont {Mayou}(2021)}]{PhysRevB.103.075415}%
  \BibitemOpen
  \bibfield  {author} {\bibinfo {author} {\bibfnamefont {G.}~\bibnamefont {Bouzerar}}\ and\ \bibinfo {author} {\bibfnamefont {D.}~\bibnamefont {Mayou}},\ }\bibfield  {title} {\bibinfo {title} {Quantum transport in flat bands and supermetallicity},\ }\href {https://doi.org/10.1103/PhysRevB.103.075415} {\bibfield  {journal} {\bibinfo  {journal} {Phys. Rev. B}\ }\textbf {\bibinfo {volume} {103}},\ \bibinfo {pages} {075415} (\bibinfo {year} {2021})}\BibitemShut {NoStop}%
\bibitem [{\citenamefont {Huhtinen}\ and\ \citenamefont {T\"orm\"a}(2023)}]{PhysRevB.108.155108}%
  \BibitemOpen
  \bibfield  {author} {\bibinfo {author} {\bibfnamefont {K.-E.}\ \bibnamefont {Huhtinen}}\ and\ \bibinfo {author} {\bibfnamefont {P.}~\bibnamefont {T\"orm\"a}},\ }\bibfield  {title} {\bibinfo {title} {Conductivity in flat bands from the kubo-greenwood formula},\ }\href {https://doi.org/10.1103/PhysRevB.108.155108} {\bibfield  {journal} {\bibinfo  {journal} {Phys. Rev. B}\ }\textbf {\bibinfo {volume} {108}},\ \bibinfo {pages} {155108} (\bibinfo {year} {2023})}\BibitemShut {NoStop}%
\bibitem [{\citenamefont {Leykam}\ \emph {et~al.}(2013)\citenamefont {Leykam}, \citenamefont {Flach}, \citenamefont {Bahat-Treidel},\ and\ \citenamefont {Desyatnikov}}]{PhysRevB.88.224203}%
  \BibitemOpen
  \bibfield  {author} {\bibinfo {author} {\bibfnamefont {D.}~\bibnamefont {Leykam}}, \bibinfo {author} {\bibfnamefont {S.}~\bibnamefont {Flach}}, \bibinfo {author} {\bibfnamefont {O.}~\bibnamefont {Bahat-Treidel}},\ and\ \bibinfo {author} {\bibfnamefont {A.~S.}\ \bibnamefont {Desyatnikov}},\ }\bibfield  {title} {\bibinfo {title} {Flat band states: Disorder and nonlinearity},\ }\href {https://doi.org/10.1103/PhysRevB.88.224203} {\bibfield  {journal} {\bibinfo  {journal} {Phys. Rev. B}\ }\textbf {\bibinfo {volume} {88}},\ \bibinfo {pages} {224203} (\bibinfo {year} {2013})}\BibitemShut {NoStop}%
\bibitem [{\citenamefont {Wilson}\ \emph {et~al.}(2020)\citenamefont {Wilson}, \citenamefont {Fu}, \citenamefont {Das~Sarma},\ and\ \citenamefont {Pixley}}]{PhysRevResearch.2.023325}%
  \BibitemOpen
  \bibfield  {author} {\bibinfo {author} {\bibfnamefont {J.~H.}\ \bibnamefont {Wilson}}, \bibinfo {author} {\bibfnamefont {Y.}~\bibnamefont {Fu}}, \bibinfo {author} {\bibfnamefont {S.}~\bibnamefont {Das~Sarma}},\ and\ \bibinfo {author} {\bibfnamefont {J.~H.}\ \bibnamefont {Pixley}},\ }\bibfield  {title} {\bibinfo {title} {Disorder in twisted bilayer graphene},\ }\href {https://doi.org/10.1103/PhysRevResearch.2.023325} {\bibfield  {journal} {\bibinfo  {journal} {Phys. Rev. Res.}\ }\textbf {\bibinfo {volume} {2}},\ \bibinfo {pages} {023325} (\bibinfo {year} {2020})}\BibitemShut {NoStop}%
\bibitem [{\citenamefont {Kazmierczak}\ \emph {et~al.}(2021)\citenamefont {Kazmierczak}, \citenamefont {Van~Winkle}, \citenamefont {Ophus}, \citenamefont {Bustillo}, \citenamefont {Carr}, \citenamefont {Brown}, \citenamefont {Ciston}, \citenamefont {Taniguchi}, \citenamefont {Watanabe},\ and\ \citenamefont {Bediako}}]{Kazmierczak2021ER}%
  \BibitemOpen
  \bibfield  {author} {\bibinfo {author} {\bibfnamefont {N.~P.}\ \bibnamefont {Kazmierczak}}, \bibinfo {author} {\bibfnamefont {M.}~\bibnamefont {Van~Winkle}}, \bibinfo {author} {\bibfnamefont {C.}~\bibnamefont {Ophus}}, \bibinfo {author} {\bibfnamefont {K.~C.}\ \bibnamefont {Bustillo}}, \bibinfo {author} {\bibfnamefont {S.}~\bibnamefont {Carr}}, \bibinfo {author} {\bibfnamefont {H.~G.}\ \bibnamefont {Brown}}, \bibinfo {author} {\bibfnamefont {J.}~\bibnamefont {Ciston}}, \bibinfo {author} {\bibfnamefont {T.}~\bibnamefont {Taniguchi}}, \bibinfo {author} {\bibfnamefont {K.}~\bibnamefont {Watanabe}},\ and\ \bibinfo {author} {\bibfnamefont {D.~K.}\ \bibnamefont {Bediako}},\ }\bibfield  {title} {\bibinfo {title} {Strain fields in twisted bilayer graphene},\ }\href {https://doi.org/10.1038/s41563-021-00973-w} {\bibfield  {journal} {\bibinfo  {journal} {Nat. Mater.}\ }\textbf {\bibinfo {volume} {20}},\ \bibinfo {pages} {956} (\bibinfo {year} {2021})}\BibitemShut {NoStop}%
\bibitem [{\citenamefont {de~Jong}\ \emph {et~al.}(2022)\citenamefont {de~Jong}, \citenamefont {Benschop}, \citenamefont {Chen}, \citenamefont {Krasovskii}, \citenamefont {de~Dood}, \citenamefont {Tromp}, \citenamefont {Allan},\ and\ \citenamefont {van~der Molen}}]{deJong2022}%
  \BibitemOpen
  \bibfield  {author} {\bibinfo {author} {\bibfnamefont {T.~A.}\ \bibnamefont {de~Jong}}, \bibinfo {author} {\bibfnamefont {T.}~\bibnamefont {Benschop}}, \bibinfo {author} {\bibfnamefont {X.}~\bibnamefont {Chen}}, \bibinfo {author} {\bibfnamefont {E.~E.}\ \bibnamefont {Krasovskii}}, \bibinfo {author} {\bibfnamefont {M.~J.~A.}\ \bibnamefont {de~Dood}}, \bibinfo {author} {\bibfnamefont {R.~M.}\ \bibnamefont {Tromp}}, \bibinfo {author} {\bibfnamefont {M.~P.}\ \bibnamefont {Allan}},\ and\ \bibinfo {author} {\bibfnamefont {S.~J.}\ \bibnamefont {van~der Molen}},\ }\bibfield  {title} {\bibinfo {title} {Imaging moir{\'e} deformation and dynamics in twisted bilayer graphene},\ }\href {https://doi.org/10.1038/s41467-021-27646-1} {\bibfield  {journal} {\bibinfo  {journal} {Nat. Commun.}\ }\textbf {\bibinfo {volume} {13}},\ \bibinfo {pages} {70} (\bibinfo {year} {2022})}\BibitemShut {NoStop}%
\bibitem [{\citenamefont {Thouless}\ \emph {et~al.}(1982)\citenamefont {Thouless}, \citenamefont {Kohmoto}, \citenamefont {Nightingale},\ and\ \citenamefont {Den~Nijs}}]{Thouless1982Aug}%
  \BibitemOpen
  \bibfield  {author} {\bibinfo {author} {\bibfnamefont {D.~J.}\ \bibnamefont {Thouless}}, \bibinfo {author} {\bibfnamefont {M.}~\bibnamefont {Kohmoto}}, \bibinfo {author} {\bibfnamefont {M.~P.}\ \bibnamefont {Nightingale}},\ and\ \bibinfo {author} {\bibfnamefont {M.}~\bibnamefont {Den~Nijs}},\ }\bibfield  {title} {\bibinfo {title} {{Quantized Hall Conductance in a Two-Dimensional Periodic Potential}},\ }\href {https://doi.org/10.1103/PhysRevLett.49.405} {\bibfield  {journal} {\bibinfo  {journal} {Phys. Rev. Lett.}\ }\textbf {\bibinfo {volume} {49}},\ \bibinfo {pages} {405} (\bibinfo {year} {1982})}\BibitemShut {NoStop}%
\bibitem [{\citenamefont {Victor}(1984)}]{Victor1984Mar}%
  \BibitemOpen
  \bibfield  {author} {\bibinfo {author} {\bibfnamefont {B.~M.}\ \bibnamefont {Victor}},\ }\bibfield  {title} {\bibinfo {title} {{Quantal phase factors accompanying adiabatic changes}},\ }\href {https://doi.org/10.1098/rspa.1984.0023} {\bibfield  {journal} {\bibinfo  {journal} {Proc. R. Soc. Lond. A.}\ }\textbf {\bibinfo {volume} {392}},\ \bibinfo {pages} {45} (\bibinfo {year} {1984})}\BibitemShut {NoStop}%
\bibitem [{\citenamefont {Kane}\ and\ \citenamefont {Mele}(2005)}]{Kane2005Sep}%
  \BibitemOpen
  \bibfield  {author} {\bibinfo {author} {\bibfnamefont {C.~L.}\ \bibnamefont {Kane}}\ and\ \bibinfo {author} {\bibfnamefont {E.~J.}\ \bibnamefont {Mele}},\ }\bibfield  {title} {\bibinfo {title} {{${Z}_{2}$ Topological Order and the Quantum Spin Hall Effect}},\ }\href {https://doi.org/10.1103/PhysRevLett.95.146802} {\bibfield  {journal} {\bibinfo  {journal} {Phys. Rev. Lett.}\ }\textbf {\bibinfo {volume} {95}},\ \bibinfo {pages} {146802} (\bibinfo {year} {2005})}\BibitemShut {NoStop}%
\bibitem [{\citenamefont {Haldane}(1988)}]{Haldane1988Oct}%
  \BibitemOpen
  \bibfield  {author} {\bibinfo {author} {\bibfnamefont {F.~D.~M.}\ \bibnamefont {Haldane}},\ }\bibfield  {title} {\bibinfo {title} {{Model for a Quantum Hall Effect without Landau Levels: Condensed-Matter Realization of the "Parity Anomaly"}},\ }\href {https://doi.org/10.1103/PhysRevLett.61.2015} {\bibfield  {journal} {\bibinfo  {journal} {Phys. Rev. Lett.}\ }\textbf {\bibinfo {volume} {61}},\ \bibinfo {pages} {2015} (\bibinfo {year} {1988})}\BibitemShut {NoStop}%
\bibitem [{\citenamefont {Provost}\ and\ \citenamefont {Vallee}(1980)}]{Provost1980Sep}%
  \BibitemOpen
  \bibfield  {author} {\bibinfo {author} {\bibfnamefont {J.~P.}\ \bibnamefont {Provost}}\ and\ \bibinfo {author} {\bibfnamefont {G.}~\bibnamefont {Vallee}},\ }\bibfield  {title} {\bibinfo {title} {{Riemannian structure on manifolds of quantum states}},\ }\href {https://doi.org/10.1007/BF02193559} {\bibfield  {journal} {\bibinfo  {journal} {Commun. Math. Phys.}\ }\textbf {\bibinfo {volume} {76}},\ \bibinfo {pages} {289} (\bibinfo {year} {1980})}\BibitemShut {NoStop}%
\bibitem [{\citenamefont {T{\ifmmode\ddot{o}\else\"{o}\fi}rm{\ifmmode\ddot{a}\else\"{a}\fi}}(2023)}]{TormaEssayQM}%
  \BibitemOpen
  \bibfield  {author} {\bibinfo {author} {\bibfnamefont {P.}~\bibnamefont {T{\ifmmode\ddot{o}\else\"{o}\fi}rm{\ifmmode\ddot{a}\else\"{a}\fi}}},\ }\bibfield  {title} {\bibinfo {title} {{Essay: Where Can Quantum Geometry Lead Us?}},\ }\href {https://doi.org/10.1103/PhysRevLett.131.240001} {\bibfield  {journal} {\bibinfo  {journal} {Phys. Rev. Lett.}\ }\textbf {\bibinfo {volume} {131}},\ \bibinfo {pages} {240001} (\bibinfo {year} {2023})}\BibitemShut {NoStop}%
\bibitem [{\citenamefont {Komissarov}\ \emph {et~al.}(2023)\citenamefont {Komissarov}, \citenamefont {Holder},\ and\ \citenamefont {Queiroz}}]{Queiroz1}%
  \BibitemOpen
  \bibfield  {author} {\bibinfo {author} {\bibfnamefont {I.}~\bibnamefont {Komissarov}}, \bibinfo {author} {\bibfnamefont {T.}~\bibnamefont {Holder}},\ and\ \bibinfo {author} {\bibfnamefont {R.}~\bibnamefont {Queiroz}},\ }\bibfield  {title} {\bibinfo {title} {{The quantum geometric origin of capacitance in insulators}},\ }\bibfield  {journal} {\bibinfo  {journal} {arXiv}\ }\href {https://doi.org/10.48550/arXiv.2306.08035} {10.48550/arXiv.2306.08035} (\bibinfo {year} {2023}),\ \Eprint {https://arxiv.org/abs/2306.08035} {2306.08035} \BibitemShut {NoStop}%
\bibitem [{\citenamefont {Verma}\ and\ \citenamefont {Queiroz}(2024)}]{Queiroz2}%
  \BibitemOpen
  \bibfield  {author} {\bibinfo {author} {\bibfnamefont {N.}~\bibnamefont {Verma}}\ and\ \bibinfo {author} {\bibfnamefont {R.}~\bibnamefont {Queiroz}},\ }\bibfield  {title} {\bibinfo {title} {{Instantaneous Response and Quantum Geometry of Insulators}},\ }\bibfield  {journal} {\bibinfo  {journal} {arXiv}\ }\href {https://doi.org/10.48550/arXiv.2403.07052} {10.48550/arXiv.2403.07052} (\bibinfo {year} {2024}),\ \Eprint {https://arxiv.org/abs/2403.07052} {2403.07052} \BibitemShut {NoStop}%
\bibitem [{\citenamefont {Pi{\ifmmode\acute{e}\else\'{e}\fi}chon}\ \emph {et~al.}(2016)\citenamefont {Pi{\ifmmode\acute{e}\else\'{e}\fi}chon}, \citenamefont {Raoux}, \citenamefont {Fuchs},\ and\ \citenamefont {Montambaux}}]{orbitalsusceptivilitywithoutberrycurvature}%
  \BibitemOpen
  \bibfield  {author} {\bibinfo {author} {\bibfnamefont {F.}~\bibnamefont {Pi{\ifmmode\acute{e}\else\'{e}\fi}chon}}, \bibinfo {author} {\bibfnamefont {A.}~\bibnamefont {Raoux}}, \bibinfo {author} {\bibfnamefont {J.-N.}\ \bibnamefont {Fuchs}},\ and\ \bibinfo {author} {\bibfnamefont {G.}~\bibnamefont {Montambaux}},\ }\bibfield  {title} {\bibinfo {title} {{Geometric orbital susceptibility: Quantum metric without Berry curvature}},\ }\href {https://doi.org/10.1103/PhysRevB.94.134423} {\bibfield  {journal} {\bibinfo  {journal} {Phys. Rev. B}\ }\textbf {\bibinfo {volume} {94}},\ \bibinfo {pages} {134423} (\bibinfo {year} {2016})}\BibitemShut {NoStop}%
\bibitem [{\citenamefont {Kruchkov}\ and\ \citenamefont {Ryu}(2023)}]{Kruchkov2023Dec}%
  \BibitemOpen
  \bibfield  {author} {\bibinfo {author} {\bibfnamefont {A.}~\bibnamefont {Kruchkov}}\ and\ \bibinfo {author} {\bibfnamefont {S.}~\bibnamefont {Ryu}},\ }\bibfield  {title} {\bibinfo {title} {{Spectral sum rules reflect topological and quantum-geometric invariants}},\ }\bibfield  {journal} {\bibinfo  {journal} {arXiv}\ }\href {https://doi.org/10.48550/arXiv.2312.17318} {10.48550/arXiv.2312.17318} (\bibinfo {year} {2023}),\ \Eprint {https://arxiv.org/abs/2312.17318} {2312.17318} \BibitemShut {NoStop}%
\bibitem [{\citenamefont {Gao}\ \emph {et~al.}(2023)\citenamefont {Gao}, \citenamefont {Liu}, \citenamefont {Qiu}, \citenamefont {Ghosh}, \citenamefont {Trevisan}, \citenamefont {Onishi}, \citenamefont {Hu}, \citenamefont {Qian}, \citenamefont {Tien}, \citenamefont {Chen}, \citenamefont {Huang}, \citenamefont {B{\ifmmode\acute{e}\else\'{e}\fi}rub{\ifmmode\acute{e}\else\'{e}\fi}}, \citenamefont {Li}, \citenamefont {Tzschaschel}, \citenamefont {Dinh}, \citenamefont {Sun}, \citenamefont {Ho}, \citenamefont {Lien}, \citenamefont {Singh}, \citenamefont {Watanabe}, \citenamefont {Taniguchi}, \citenamefont {Bell}, \citenamefont {Lin}, \citenamefont {Chang}, \citenamefont {Du}, \citenamefont {Bansil}, \citenamefont {Fu}, \citenamefont {Ni}, \citenamefont {Orth}, \citenamefont {Ma},\ and\ \citenamefont {Xu}}]{Gao2023Jun}%
  \BibitemOpen
  \bibfield  {author} {\bibinfo {author} {\bibfnamefont {A.}~\bibnamefont {Gao}}, \bibinfo {author} {\bibfnamefont {Y.-F.}\ \bibnamefont {Liu}}, \bibinfo {author} {\bibfnamefont {J.-X.}\ \bibnamefont {Qiu}}, \bibinfo {author} {\bibfnamefont {B.}~\bibnamefont {Ghosh}}, \bibinfo {author} {\bibfnamefont {T.~V.}\ \bibnamefont {Trevisan}}, \bibinfo {author} {\bibfnamefont {Y.}~\bibnamefont {Onishi}}, \bibinfo {author} {\bibfnamefont {C.}~\bibnamefont {Hu}}, \bibinfo {author} {\bibfnamefont {T.}~\bibnamefont {Qian}}, \bibinfo {author} {\bibfnamefont {H.-J.}\ \bibnamefont {Tien}}, \bibinfo {author} {\bibfnamefont {S.-W.}\ \bibnamefont {Chen}}, \bibinfo {author} {\bibfnamefont {M.}~\bibnamefont {Huang}}, \bibinfo {author} {\bibfnamefont {D.}~\bibnamefont {B{\ifmmode\acute{e}\else\'{e}\fi}rub{\ifmmode\acute{e}\else\'{e}\fi}}}, \bibinfo {author} {\bibfnamefont {H.}~\bibnamefont {Li}}, \bibinfo {author} {\bibfnamefont {C.}~\bibnamefont {Tzschaschel}}, \bibinfo {author} {\bibfnamefont {T.}~\bibnamefont {Dinh}}, \bibinfo
  {author} {\bibfnamefont {Z.}~\bibnamefont {Sun}}, \bibinfo {author} {\bibfnamefont {S.-C.}\ \bibnamefont {Ho}}, \bibinfo {author} {\bibfnamefont {S.-W.}\ \bibnamefont {Lien}}, \bibinfo {author} {\bibfnamefont {B.}~\bibnamefont {Singh}}, \bibinfo {author} {\bibfnamefont {K.}~\bibnamefont {Watanabe}}, \bibinfo {author} {\bibfnamefont {T.}~\bibnamefont {Taniguchi}}, \bibinfo {author} {\bibfnamefont {D.~C.}\ \bibnamefont {Bell}}, \bibinfo {author} {\bibfnamefont {H.}~\bibnamefont {Lin}}, \bibinfo {author} {\bibfnamefont {T.-R.}\ \bibnamefont {Chang}}, \bibinfo {author} {\bibfnamefont {C.~R.}\ \bibnamefont {Du}}, \bibinfo {author} {\bibfnamefont {A.}~\bibnamefont {Bansil}}, \bibinfo {author} {\bibfnamefont {L.}~\bibnamefont {Fu}}, \bibinfo {author} {\bibfnamefont {N.}~\bibnamefont {Ni}}, \bibinfo {author} {\bibfnamefont {P.~P.}\ \bibnamefont {Orth}}, \bibinfo {author} {\bibfnamefont {Q.}~\bibnamefont {Ma}},\ and\ \bibinfo {author} {\bibfnamefont {S.-Y.}\ \bibnamefont {Xu}},\ }\bibfield  {title} {\bibinfo {title}
  {{Quantum metric nonlinear Hall effect in a topological antiferromagnetic heterostructure}},\ }\href {https://doi.org/10.1126/science.adf1506} {\bibfield  {journal} {\bibinfo  {journal} {Science}\ }\textbf {\bibinfo {volume} {381}},\ \bibinfo {pages} {181} (\bibinfo {year} {2023})}\BibitemShut {NoStop}%
\bibitem [{\citenamefont {Gao}\ and\ \citenamefont {Xiao}(2019)}]{Gao2019Jun}%
  \BibitemOpen
  \bibfield  {author} {\bibinfo {author} {\bibfnamefont {Y.}~\bibnamefont {Gao}}\ and\ \bibinfo {author} {\bibfnamefont {D.}~\bibnamefont {Xiao}},\ }\bibfield  {title} {\bibinfo {title} {{Nonreciprocal Directional Dichroism Induced by the Quantum Metric Dipole}},\ }\href {https://doi.org/10.1103/PhysRevLett.122.227402} {\bibfield  {journal} {\bibinfo  {journal} {Phys. Rev. Lett.}\ }\textbf {\bibinfo {volume} {122}},\ \bibinfo {pages} {227402} (\bibinfo {year} {2019})}\BibitemShut {NoStop}%
\bibitem [{\citenamefont {Wang}\ \emph {et~al.}(2021)\citenamefont {Wang}, \citenamefont {Gao},\ and\ \citenamefont {Xiao}}]{Wang2021Dec}%
  \BibitemOpen
  \bibfield  {author} {\bibinfo {author} {\bibfnamefont {C.}~\bibnamefont {Wang}}, \bibinfo {author} {\bibfnamefont {Y.}~\bibnamefont {Gao}},\ and\ \bibinfo {author} {\bibfnamefont {D.}~\bibnamefont {Xiao}},\ }\bibfield  {title} {\bibinfo {title} {{Intrinsic Nonlinear Hall Effect in Antiferromagnetic Tetragonal CuMnAs}},\ }\href {https://doi.org/10.1103/PhysRevLett.127.277201} {\bibfield  {journal} {\bibinfo  {journal} {Phys. Rev. Lett.}\ }\textbf {\bibinfo {volume} {127}},\ \bibinfo {pages} {277201} (\bibinfo {year} {2021})}\BibitemShut {NoStop}%
\bibitem [{\citenamefont {Das}\ \emph {et~al.}(2023)\citenamefont {Das}, \citenamefont {Lahiri}, \citenamefont {Atencia}, \citenamefont {Culcer},\ and\ \citenamefont {Agarwal}}]{Das2023Nov}%
  \BibitemOpen
  \bibfield  {author} {\bibinfo {author} {\bibfnamefont {K.}~\bibnamefont {Das}}, \bibinfo {author} {\bibfnamefont {S.}~\bibnamefont {Lahiri}}, \bibinfo {author} {\bibfnamefont {R.~B.}\ \bibnamefont {Atencia}}, \bibinfo {author} {\bibfnamefont {D.}~\bibnamefont {Culcer}},\ and\ \bibinfo {author} {\bibfnamefont {A.}~\bibnamefont {Agarwal}},\ }\bibfield  {title} {\bibinfo {title} {{Intrinsic nonlinear conductivities induced by the quantum metric}},\ }\href {https://doi.org/10.1103/PhysRevB.108.L201405} {\bibfield  {journal} {\bibinfo  {journal} {Phys. Rev. B}\ }\textbf {\bibinfo {volume} {108}},\ \bibinfo {pages} {L201405} (\bibinfo {year} {2023})}\BibitemShut {NoStop}%
\bibitem [{\citenamefont {Kruchkov}(2023)}]{Kruchkov2023Jun}%
  \BibitemOpen
  \bibfield  {author} {\bibinfo {author} {\bibfnamefont {A.}~\bibnamefont {Kruchkov}},\ }\bibfield  {title} {\bibinfo {title} {{Quantum transport anomalies in dispersionless quantum states}},\ }\href {https://doi.org/10.1103/PhysRevB.107.L241102} {\bibfield  {journal} {\bibinfo  {journal} {Phys. Rev. B}\ }\textbf {\bibinfo {volume} {107}},\ \bibinfo {pages} {L241102} (\bibinfo {year} {2023})}\BibitemShut {NoStop}%
\bibitem [{\citenamefont {Mera}\ and\ \citenamefont {Mitscherling}(2022)}]{Mera2022Oct}%
  \BibitemOpen
  \bibfield  {author} {\bibinfo {author} {\bibfnamefont {B.}~\bibnamefont {Mera}}\ and\ \bibinfo {author} {\bibfnamefont {J.}~\bibnamefont {Mitscherling}},\ }\bibfield  {title} {\bibinfo {title} {{Nontrivial quantum geometry of degenerate flat bands}},\ }\href {https://doi.org/10.1103/PhysRevB.106.165133} {\bibfield  {journal} {\bibinfo  {journal} {Phys. Rev. B}\ }\textbf {\bibinfo {volume} {106}},\ \bibinfo {pages} {165133} (\bibinfo {year} {2022})}\BibitemShut {NoStop}%
\bibitem [{\citenamefont {Rhim}\ \emph {et~al.}(2020)\citenamefont {Rhim}, \citenamefont {Kim},\ and\ \citenamefont {Yang}}]{Rhim2020Aug}%
  \BibitemOpen
  \bibfield  {author} {\bibinfo {author} {\bibfnamefont {J.-W.}\ \bibnamefont {Rhim}}, \bibinfo {author} {\bibfnamefont {K.}~\bibnamefont {Kim}},\ and\ \bibinfo {author} {\bibfnamefont {B.-J.}\ \bibnamefont {Yang}},\ }\bibfield  {title} {\bibinfo {title} {{Quantum distance and anomalous Landau levels of flat bands}},\ }\href {https://doi.org/10.1038/s41586-020-2540-1} {\bibfield  {journal} {\bibinfo  {journal} {Nature}\ }\textbf {\bibinfo {volume} {584}},\ \bibinfo {pages} {59} (\bibinfo {year} {2020})}\BibitemShut {NoStop}%
\bibitem [{\citenamefont {Peotta}\ and\ \citenamefont {T{\ifmmode\ddot{o}\else\"{o}\fi}rm{\ifmmode\ddot{a}\else\"{a}\fi}}(2015)}]{Peotta2015Nov}%
  \BibitemOpen
  \bibfield  {author} {\bibinfo {author} {\bibfnamefont {S.}~\bibnamefont {Peotta}}\ and\ \bibinfo {author} {\bibfnamefont {P.}~\bibnamefont {T{\ifmmode\ddot{o}\else\"{o}\fi}rm{\ifmmode\ddot{a}\else\"{a}\fi}}},\ }\bibfield  {title} {\bibinfo {title} {{Superfluidity in topologically nontrivial flat bands}},\ }\href {https://doi.org/10.1038/ncomms9944} {\bibfield  {journal} {\bibinfo  {journal} {Nat. Commun.}\ }\textbf {\bibinfo {volume} {6}},\ \bibinfo {pages} {1} (\bibinfo {year} {2015})}\BibitemShut {NoStop}%
\bibitem [{\citenamefont {Julku}\ \emph {et~al.}(2020)\citenamefont {Julku}, \citenamefont {Peltonen}, \citenamefont {Liang}, \citenamefont {Heikkil{\ifmmode\ddot{a}\else\"{a}\fi}},\ and\ \citenamefont {T{\ifmmode\ddot{o}\else\"{o}\fi}rm{\ifmmode\ddot{a}\else\"{a}\fi}}}]{Julku2020Feb}%
  \BibitemOpen
  \bibfield  {author} {\bibinfo {author} {\bibfnamefont {A.}~\bibnamefont {Julku}}, \bibinfo {author} {\bibfnamefont {T.~J.}\ \bibnamefont {Peltonen}}, \bibinfo {author} {\bibfnamefont {L.}~\bibnamefont {Liang}}, \bibinfo {author} {\bibfnamefont {T.~T.}\ \bibnamefont {Heikkil{\ifmmode\ddot{a}\else\"{a}\fi}}},\ and\ \bibinfo {author} {\bibfnamefont {P.}~\bibnamefont {T{\ifmmode\ddot{o}\else\"{o}\fi}rm{\ifmmode\ddot{a}\else\"{a}\fi}}},\ }\bibfield  {title} {\bibinfo {title} {{Superfluid weight and Berezinskii-Kosterlitz-Thouless transition temperature of twisted bilayer graphene}},\ }\href {https://doi.org/10.1103/PhysRevB.101.060505} {\bibfield  {journal} {\bibinfo  {journal} {Phys. Rev. B}\ }\textbf {\bibinfo {volume} {101}},\ \bibinfo {pages} {060505} (\bibinfo {year} {2020})}\BibitemShut {NoStop}%
\bibitem [{\citenamefont {Peri}\ \emph {et~al.}(2021)\citenamefont {Peri}, \citenamefont {Song}, \citenamefont {Bernevig},\ and\ \citenamefont {Huber}}]{Peri2021Jan}%
  \BibitemOpen
  \bibfield  {author} {\bibinfo {author} {\bibfnamefont {V.}~\bibnamefont {Peri}}, \bibinfo {author} {\bibfnamefont {Z.-D.}\ \bibnamefont {Song}}, \bibinfo {author} {\bibfnamefont {B.~A.}\ \bibnamefont {Bernevig}},\ and\ \bibinfo {author} {\bibfnamefont {S.~D.}\ \bibnamefont {Huber}},\ }\bibfield  {title} {\bibinfo {title} {{Fragile Topology and Flat-Band Superconductivity in the Strong-Coupling Regime}},\ }\href {https://doi.org/10.1103/PhysRevLett.126.027002} {\bibfield  {journal} {\bibinfo  {journal} {Phys. Rev. Lett.}\ }\textbf {\bibinfo {volume} {126}},\ \bibinfo {pages} {027002} (\bibinfo {year} {2021})}\BibitemShut {NoStop}%
\bibitem [{\citenamefont {Rossi}(2021)}]{Rossi2021}%
  \BibitemOpen
  \bibfield  {author} {\bibinfo {author} {\bibfnamefont {E.}~\bibnamefont {Rossi}},\ }\bibfield  {title} {\bibinfo {title} {Quantum metric and correlated states in two-dimensional systems},\ }\href {https://doi.org/https://doi.org/10.1016/j.cossms.2021.100952} {\bibfield  {journal} {\bibinfo  {journal} {Curr. Opin. Solid State Mater. Sci.}\ }\textbf {\bibinfo {volume} {25}},\ \bibinfo {pages} {100952} (\bibinfo {year} {2021})}\BibitemShut {NoStop}%
\bibitem [{\citenamefont {T{\"o}rm{\"a}}\ \emph {et~al.}(2022)\citenamefont {T{\"o}rm{\"a}}, \citenamefont {Peotta},\ and\ \citenamefont {Bernevig}}]{Torma2022}%
  \BibitemOpen
  \bibfield  {author} {\bibinfo {author} {\bibfnamefont {P.}~\bibnamefont {T{\"o}rm{\"a}}}, \bibinfo {author} {\bibfnamefont {S.}~\bibnamefont {Peotta}},\ and\ \bibinfo {author} {\bibfnamefont {B.~A.}\ \bibnamefont {Bernevig}},\ }\bibfield  {title} {\bibinfo {title} {Superconductivity, superfluidity and quantum geometry in twisted multilayer systems},\ }\href {https://doi.org/10.1038/s42254-022-00466-y} {\bibfield  {journal} {\bibinfo  {journal} {Nat. Rev. Phys.}\ }\textbf {\bibinfo {volume} {4}},\ \bibinfo {pages} {528} (\bibinfo {year} {2022})}\BibitemShut {NoStop}%
\bibitem [{\citenamefont {Kohn}(1964)}]{theoryofinsulatingKohn}%
  \BibitemOpen
  \bibfield  {author} {\bibinfo {author} {\bibfnamefont {W.}~\bibnamefont {Kohn}},\ }\bibfield  {title} {\bibinfo {title} {{Theory of the Insulating State}},\ }\href {https://doi.org/10.1103/PhysRev.133.A171} {\bibfield  {journal} {\bibinfo  {journal} {Phys. Rev.}\ }\textbf {\bibinfo {volume} {133}},\ \bibinfo {pages} {A171} (\bibinfo {year} {1964})}\BibitemShut {NoStop}%
\bibitem [{\citenamefont {Resta}(2018)}]{ModerntheoryinsulatingResta}%
  \BibitemOpen
  \bibfield  {author} {\bibinfo {author} {\bibfnamefont {R.}~\bibnamefont {Resta}},\ }\bibfield  {title} {\bibinfo {title} {{Theory of the insulating state}},\ }\href {https://doi.org/10.1393/ncr/i2018-10151-1} {\bibfield  {journal} {\bibinfo  {journal} {Riv. Nuovo Cimento}\ }\textbf {\bibinfo {volume} {41}},\ \bibinfo {pages} {463} (\bibinfo {year} {2018})}\BibitemShut {NoStop}%
\bibitem [{\citenamefont {Marzari}\ \emph {et~al.}(2012)\citenamefont {Marzari}, \citenamefont {Mostofi}, \citenamefont {Yates}, \citenamefont {Souza},\ and\ \citenamefont {Vanderbilt}}]{MaximallylocalizedWannierfunctions}%
  \BibitemOpen
  \bibfield  {author} {\bibinfo {author} {\bibfnamefont {N.}~\bibnamefont {Marzari}}, \bibinfo {author} {\bibfnamefont {A.~A.}\ \bibnamefont {Mostofi}}, \bibinfo {author} {\bibfnamefont {J.~R.}\ \bibnamefont {Yates}}, \bibinfo {author} {\bibfnamefont {I.}~\bibnamefont {Souza}},\ and\ \bibinfo {author} {\bibfnamefont {D.}~\bibnamefont {Vanderbilt}},\ }\bibfield  {title} {\bibinfo {title} {{Maximally localized Wannier functions: Theory and applications}},\ }\href {https://doi.org/10.1103/RevModPhys.84.1419} {\bibfield  {journal} {\bibinfo  {journal} {Rev. Mod. Phys.}\ }\textbf {\bibinfo {volume} {84}},\ \bibinfo {pages} {1419} (\bibinfo {year} {2012})}\BibitemShut {NoStop}%
\bibitem [{\citenamefont {Marrazzo}\ and\ \citenamefont {Resta}(2019)}]{Localtheoryofinsulatingstate}%
  \BibitemOpen
  \bibfield  {author} {\bibinfo {author} {\bibfnamefont {A.}~\bibnamefont {Marrazzo}}\ and\ \bibinfo {author} {\bibfnamefont {R.}~\bibnamefont {Resta}},\ }\bibfield  {title} {\bibinfo {title} {{Local Theory of the Insulating State}},\ }\href {https://doi.org/10.1103/PhysRevLett.122.166602} {\bibfield  {journal} {\bibinfo  {journal} {Phys. Rev. Lett.}\ }\textbf {\bibinfo {volume} {122}},\ \bibinfo {pages} {166602} (\bibinfo {year} {2019})}\BibitemShut {NoStop}%
\bibitem [{\citenamefont {Fan}\ \emph {et~al.}(2021)\citenamefont {Fan}, \citenamefont {Garcia}, \citenamefont {Cummings}, \citenamefont {Barrios-Vargas}, \citenamefont {Panhans}, \citenamefont {Harju}, \citenamefont {Ortmann},\ and\ \citenamefont {Roche}}]{FAN20211}%
  \BibitemOpen
  \bibfield  {author} {\bibinfo {author} {\bibfnamefont {Z.}~\bibnamefont {Fan}}, \bibinfo {author} {\bibfnamefont {J.~H.}\ \bibnamefont {Garcia}}, \bibinfo {author} {\bibfnamefont {A.~W.}\ \bibnamefont {Cummings}}, \bibinfo {author} {\bibfnamefont {J.~E.}\ \bibnamefont {Barrios-Vargas}}, \bibinfo {author} {\bibfnamefont {M.}~\bibnamefont {Panhans}}, \bibinfo {author} {\bibfnamefont {A.}~\bibnamefont {Harju}}, \bibinfo {author} {\bibfnamefont {F.}~\bibnamefont {Ortmann}},\ and\ \bibinfo {author} {\bibfnamefont {S.}~\bibnamefont {Roche}},\ }\bibfield  {title} {\bibinfo {title} {Linear scaling quantum transport methodologies},\ }\href {https://doi.org/https://doi.org/10.1016/j.physrep.2020.12.001} {\bibfield  {journal} {\bibinfo  {journal} {Phys. Rep.}\ }\textbf {\bibinfo {volume} {903}},\ \bibinfo {pages} {1} (\bibinfo {year} {2021})}\BibitemShut {NoStop}%
\bibitem [{\citenamefont {Souza}\ \emph {et~al.}(2000)\citenamefont {Souza}, \citenamefont {Wilkens},\ and\ \citenamefont {Martin}}]{SWMrule}%
  \BibitemOpen
  \bibfield  {author} {\bibinfo {author} {\bibfnamefont {I.}~\bibnamefont {Souza}}, \bibinfo {author} {\bibfnamefont {T.}~\bibnamefont {Wilkens}},\ and\ \bibinfo {author} {\bibfnamefont {R.~M.}\ \bibnamefont {Martin}},\ }\bibfield  {title} {\bibinfo {title} {{Polarization and localization in insulators: Generating function approach}},\ }\href {https://doi.org/10.1103/PhysRevB.62.1666} {\bibfield  {journal} {\bibinfo  {journal} {Phys. Rev. B}\ }\textbf {\bibinfo {volume} {62}},\ \bibinfo {pages} {1666} (\bibinfo {year} {2000})}\BibitemShut {NoStop}%
\bibitem [{\citenamefont {Jo{\ifmmode\tilde{a}\else\~{a}\fi}o}\ \emph {et~al.}(2020)\citenamefont {Jo{\ifmmode\tilde{a}\else\~{a}\fi}o}, \citenamefont {An{\dj}elkovi{\ifmmode\acute{c}\else\'{c}\fi}}, \citenamefont {Covaci}, \citenamefont {Rappoport}, \citenamefont {Lopes},\ and\ \citenamefont {Ferreira}}]{opticalconductivity}%
  \BibitemOpen
  \bibfield  {author} {\bibinfo {author} {\bibfnamefont {S.~M.}\ \bibnamefont {Jo{\ifmmode\tilde{a}\else\~{a}\fi}o}}, \bibinfo {author} {\bibfnamefont {M.}~\bibnamefont {An{\dj}elkovi{\ifmmode\acute{c}\else\'{c}\fi}}}, \bibinfo {author} {\bibfnamefont {L.}~\bibnamefont {Covaci}}, \bibinfo {author} {\bibfnamefont {T.~G.}\ \bibnamefont {Rappoport}}, \bibinfo {author} {\bibfnamefont {J.~M. V.~P.}\ \bibnamefont {Lopes}},\ and\ \bibinfo {author} {\bibfnamefont {A.}~\bibnamefont {Ferreira}},\ }\bibfield  {title} {\bibinfo {title} {{KITE: high-performance accurate modelling of electronic structure and response functions of large molecules, disordered crystals and heterostructures}},\ }\bibfield  {journal} {\bibinfo  {journal} {R. Soc. Open Sci.}\ }\href {https://doi.org/10.1098/rsos.191809} {10.1098/rsos.191809} (\bibinfo {year} {2020})\BibitemShut {NoStop}%
\bibitem [{\citenamefont {Nguyen}\ \emph {et~al.}(2022)\citenamefont {Nguyen}, \citenamefont {Hoang},\ and\ \citenamefont {Charlier}}]{Nguyen_2022}%
  \BibitemOpen
  \bibfield  {author} {\bibinfo {author} {\bibfnamefont {V.~H.}\ \bibnamefont {Nguyen}}, \bibinfo {author} {\bibfnamefont {T.~X.}\ \bibnamefont {Hoang}},\ and\ \bibinfo {author} {\bibfnamefont {J.-C.}\ \bibnamefont {Charlier}},\ }\bibfield  {title} {\bibinfo {title} {Electronic properties of twisted multilayer graphene},\ }\href {https://doi.org/10.1088/2515-7639/ac6c4a} {\bibfield  {journal} {\bibinfo  {journal} {J. Phys. Mater.}\ }\textbf {\bibinfo {volume} {5}},\ \bibinfo {pages} {034003} (\bibinfo {year} {2022})}\BibitemShut {NoStop}%
\bibitem [{\citenamefont {Lindsay}\ and\ \citenamefont {Broido}(2010)}]{Lindsay2010}%
  \BibitemOpen
  \bibfield  {author} {\bibinfo {author} {\bibfnamefont {L.}~\bibnamefont {Lindsay}}\ and\ \bibinfo {author} {\bibfnamefont {D.~A.}\ \bibnamefont {Broido}},\ }\bibfield  {title} {\bibinfo {title} {Optimized tersoff and brenner empirical potential parameters for lattice dynamics and phonon thermal transport in carbon nanotubes and graphene},\ }\href {https://doi.org/10.1103/PhysRevB.81.205441} {\bibfield  {journal} {\bibinfo  {journal} {Phys. Rev. B}\ }\textbf {\bibinfo {volume} {81}},\ \bibinfo {pages} {205441} (\bibinfo {year} {2010})}\BibitemShut {NoStop}%
\bibitem [{\citenamefont {Kolmogorov}\ and\ \citenamefont {Crespi}(2005)}]{Kolmogorov2005}%
  \BibitemOpen
  \bibfield  {author} {\bibinfo {author} {\bibfnamefont {A.~N.}\ \bibnamefont {Kolmogorov}}\ and\ \bibinfo {author} {\bibfnamefont {V.~H.}\ \bibnamefont {Crespi}},\ }\bibfield  {title} {\bibinfo {title} {Registry-dependent interlayer potential for graphitic systems},\ }\href {https://doi.org/10.1103/PhysRevB.71.235415} {\bibfield  {journal} {\bibinfo  {journal} {Phys. Rev. B}\ }\textbf {\bibinfo {volume} {71}},\ \bibinfo {pages} {235415} (\bibinfo {year} {2005})}\BibitemShut {NoStop}%
\bibitem [{\citenamefont {Leven}\ \emph {et~al.}(2016)\citenamefont {Leven}, \citenamefont {Maaravi}, \citenamefont {Azuri}, \citenamefont {Kronik},\ and\ \citenamefont {Hod}}]{Leven2016}%
  \BibitemOpen
  \bibfield  {author} {\bibinfo {author} {\bibfnamefont {I.}~\bibnamefont {Leven}}, \bibinfo {author} {\bibfnamefont {T.}~\bibnamefont {Maaravi}}, \bibinfo {author} {\bibfnamefont {I.}~\bibnamefont {Azuri}}, \bibinfo {author} {\bibfnamefont {L.}~\bibnamefont {Kronik}},\ and\ \bibinfo {author} {\bibfnamefont {O.}~\bibnamefont {Hod}},\ }\bibfield  {title} {\bibinfo {title} {Interlayer potential for graphene/h-bn heterostructures},\ }\href {https://doi.org/10.1021/acs.jctc.6b00147} {\bibfield  {journal} {\bibinfo  {journal} {J. Chem. Theory Comput.}\ }\textbf {\bibinfo {volume} {12}},\ \bibinfo {pages} {2896} (\bibinfo {year} {2016})}\BibitemShut {NoStop}%
\bibitem [{\citenamefont {Trambly~de Laissardière}\ \emph {et~al.}(2010)\citenamefont {Trambly~de Laissardière}, \citenamefont {Mayou},\ and\ \citenamefont {Magaud}}]{Trambly2010}%
  \BibitemOpen
  \bibfield  {author} {\bibinfo {author} {\bibfnamefont {G.}~\bibnamefont {Trambly~de Laissardière}}, \bibinfo {author} {\bibfnamefont {D.}~\bibnamefont {Mayou}},\ and\ \bibinfo {author} {\bibfnamefont {L.}~\bibnamefont {Magaud}},\ }\bibfield  {title} {\bibinfo {title} {Localization of dirac electrons in rotated graphene bilayers},\ }\href {https://doi.org/10.1021/nl902948m} {\bibfield  {journal} {\bibinfo  {journal} {Nano Lett.}\ }\textbf {\bibinfo {volume} {10}},\ \bibinfo {pages} {804} (\bibinfo {year} {2010})}\BibitemShut {NoStop}%
\bibitem [{\citenamefont {Trambly~de Laissardi\`ere}\ \emph {et~al.}(2012)\citenamefont {Trambly~de Laissardi\`ere}, \citenamefont {Mayou},\ and\ \citenamefont {Magaud}}]{Trambly2012}%
  \BibitemOpen
  \bibfield  {author} {\bibinfo {author} {\bibfnamefont {G.}~\bibnamefont {Trambly~de Laissardi\`ere}}, \bibinfo {author} {\bibfnamefont {D.}~\bibnamefont {Mayou}},\ and\ \bibinfo {author} {\bibfnamefont {L.}~\bibnamefont {Magaud}},\ }\bibfield  {title} {\bibinfo {title} {Numerical studies of confined states in rotated bilayers of graphene},\ }\href {https://doi.org/10.1103/PhysRevB.86.125413} {\bibfield  {journal} {\bibinfo  {journal} {Phys. Rev. B}\ }\textbf {\bibinfo {volume} {86}},\ \bibinfo {pages} {125413} (\bibinfo {year} {2012})}\BibitemShut {NoStop}%
\bibitem [{\citenamefont {Luis E. F. Foa~Torres}(2002)}]{FoaTorres}%
  \BibitemOpen
  \bibfield  {author} {\bibinfo {author} {\bibfnamefont {J.-C.~C.}\ \bibnamefont {Luis E. F. Foa~Torres}, \bibfnamefont {Stephan~Roche}},\ }\href@noop {} {\emph {\bibinfo {title} {Introduction to Graphene-Based Nanomaterials}}}\ (\bibinfo  {publisher} {Cambridge University Press},\ \bibinfo {year} {2002})\BibitemShut {NoStop}%
\bibitem [{\citenamefont {Tan}\ \emph {et~al.}(2007)\citenamefont {Tan}, \citenamefont {Zhang}, \citenamefont {Bolotin}, \citenamefont {Zhao}, \citenamefont {Adam}, \citenamefont {Hwang}, \citenamefont {Das~Sarma}, \citenamefont {Stormer},\ and\ \citenamefont {Kim}}]{Tan2007}%
  \BibitemOpen
  \bibfield  {author} {\bibinfo {author} {\bibfnamefont {Y.-W.}\ \bibnamefont {Tan}}, \bibinfo {author} {\bibfnamefont {Y.}~\bibnamefont {Zhang}}, \bibinfo {author} {\bibfnamefont {K.}~\bibnamefont {Bolotin}}, \bibinfo {author} {\bibfnamefont {Y.}~\bibnamefont {Zhao}}, \bibinfo {author} {\bibfnamefont {S.}~\bibnamefont {Adam}}, \bibinfo {author} {\bibfnamefont {E.~H.}\ \bibnamefont {Hwang}}, \bibinfo {author} {\bibfnamefont {S.}~\bibnamefont {Das~Sarma}}, \bibinfo {author} {\bibfnamefont {H.~L.}\ \bibnamefont {Stormer}},\ and\ \bibinfo {author} {\bibfnamefont {P.}~\bibnamefont {Kim}},\ }\bibfield  {title} {\bibinfo {title} {{Measurement of Scattering Rate and Minimum Conductivity in Graphene}},\ }\href {https://doi.org/10.1103/PhysRevLett.99.246803} {\bibfield  {journal} {\bibinfo  {journal} {Phys. Rev. Lett.}\ }\textbf {\bibinfo {volume} {99}},\ \bibinfo {pages} {246803} (\bibinfo {year} {2007})}\BibitemShut {NoStop}%
\bibitem [{\citenamefont {Wei\ss{}e}\ \emph {et~al.}(2006)\citenamefont {Wei\ss{}e}, \citenamefont {Wellein}, \citenamefont {Alvermann},\ and\ \citenamefont {Fehske}}]{WeisseKPM}%
  \BibitemOpen
  \bibfield  {author} {\bibinfo {author} {\bibfnamefont {A.}~\bibnamefont {Wei\ss{}e}}, \bibinfo {author} {\bibfnamefont {G.}~\bibnamefont {Wellein}}, \bibinfo {author} {\bibfnamefont {A.}~\bibnamefont {Alvermann}},\ and\ \bibinfo {author} {\bibfnamefont {H.}~\bibnamefont {Fehske}},\ }\bibfield  {title} {\bibinfo {title} {{The kernel polynomial method}},\ }\href {https://doi.org/10.1103/RevModPhys.78.275} {\bibfield  {journal} {\bibinfo  {journal} {Rev. Mod. Phys.}\ }\textbf {\bibinfo {volume} {78}},\ \bibinfo {pages} {275} (\bibinfo {year} {2006})}\BibitemShut {NoStop}%
\bibitem [{\citenamefont {Kubo}(1957)}]{Kubo}%
  \BibitemOpen
  \bibfield  {author} {\bibinfo {author} {\bibfnamefont {R.}~\bibnamefont {Kubo}},\ }\bibfield  {title} {\bibinfo {title} {{Statistical-Mechanical Theory of Irreversible Processes. I. General Theory and Simple Applications to Magnetic and Conduction Problems}},\ }\href {https://doi.org/10.1143/JPSJ.12.570} {\bibfield  {journal} {\bibinfo  {journal} {J. Phys. Soc. Jpn.}\ }\textbf {\bibinfo {volume} {12}},\ \bibinfo {pages} {570} (\bibinfo {year} {1957})}\BibitemShut {NoStop}%
\bibitem [{sup()}]{suppmaterial}%
  \BibitemOpen
  \href@noop {} {}\bibinfo {note} {See Supplemental Material at [URL will be inserted by publisher] for a detailed description of the converge of the Quantum metric with the number of polynomials and size of the system and short time evolution of the wavepackage.}\BibitemShut {Stop}%
\bibitem [{\citenamefont {Choi}\ and\ \citenamefont {Choi}(2018)}]{Choi2018}%
  \BibitemOpen
  \bibfield  {author} {\bibinfo {author} {\bibfnamefont {Y.~W.}\ \bibnamefont {Choi}}\ and\ \bibinfo {author} {\bibfnamefont {H.~J.}\ \bibnamefont {Choi}},\ }\bibfield  {title} {\bibinfo {title} {Strong electron-phonon coupling, electron-hole asymmetry, and nonadiabaticity in magic-angle twisted bilayer graphene},\ }\href {https://doi.org/10.1103/PhysRevB.98.241412} {\bibfield  {journal} {\bibinfo  {journal} {Phys. Rev. B}\ }\textbf {\bibinfo {volume} {98}},\ \bibinfo {pages} {241412} (\bibinfo {year} {2018})}\BibitemShut {NoStop}%
\bibitem [{\citenamefont {Gadelha}\ \emph {et~al.}(2022)\citenamefont {Gadelha}, \citenamefont {Nguyen}, \citenamefont {Neto}, \citenamefont {Santana}, \citenamefont {Raschke}, \citenamefont {Lamparski}, \citenamefont {Meunier}, \citenamefont {Charlier},\ and\ \citenamefont {Jorio}}]{Gadelha2022}%
  \BibitemOpen
  \bibfield  {author} {\bibinfo {author} {\bibfnamefont {A.~C.}\ \bibnamefont {Gadelha}}, \bibinfo {author} {\bibfnamefont {V.-H.}\ \bibnamefont {Nguyen}}, \bibinfo {author} {\bibfnamefont {E.~G.~S.}\ \bibnamefont {Neto}}, \bibinfo {author} {\bibfnamefont {F.}~\bibnamefont {Santana}}, \bibinfo {author} {\bibfnamefont {M.~B.}\ \bibnamefont {Raschke}}, \bibinfo {author} {\bibfnamefont {M.}~\bibnamefont {Lamparski}}, \bibinfo {author} {\bibfnamefont {V.}~\bibnamefont {Meunier}}, \bibinfo {author} {\bibfnamefont {J.-C.}\ \bibnamefont {Charlier}},\ and\ \bibinfo {author} {\bibfnamefont {A.}~\bibnamefont {Jorio}},\ }\bibfield  {title} {\bibinfo {title} {Electron–phonon coupling in a magic-angle twisted-bilayer graphene device from gate-dependent raman spectroscopy and atomistic modeling},\ }\href {https://doi.org/10.1021/acs.nanolett.2c00905} {\bibfield  {journal} {\bibinfo  {journal} {Nano Lett.}\ }\textbf {\bibinfo {volume} {22}},\ \bibinfo {pages} {6069} (\bibinfo {year} {2022})}\BibitemShut {NoStop}%
\bibitem [{\citenamefont {Van~Tuan}\ \emph {et~al.}(2016)\citenamefont {Van~Tuan}, \citenamefont {Ortmann}, \citenamefont {Cummings}, \citenamefont {Soriano},\ and\ \citenamefont {Roche}}]{VanTuan2016}%
  \BibitemOpen
  \bibfield  {author} {\bibinfo {author} {\bibfnamefont {D.}~\bibnamefont {Van~Tuan}}, \bibinfo {author} {\bibfnamefont {F.}~\bibnamefont {Ortmann}}, \bibinfo {author} {\bibfnamefont {A.~W.}\ \bibnamefont {Cummings}}, \bibinfo {author} {\bibfnamefont {D.}~\bibnamefont {Soriano}},\ and\ \bibinfo {author} {\bibfnamefont {S.}~\bibnamefont {Roche}},\ }\bibfield  {title} {\bibinfo {title} {Spin dynamics and relaxation in graphene dictated by electron-hole puddles},\ }\href {https://doi.org/10.1038/srep21046} {\bibfield  {journal} {\bibinfo  {journal} {Sci. Rep.}\ }\textbf {\bibinfo {volume} {6}},\ \bibinfo {pages} {21046} (\bibinfo {year} {2016})}\BibitemShut {NoStop}%
\bibitem [{\citenamefont {Lherbier}\ \emph {et~al.}(2008)\citenamefont {Lherbier}, \citenamefont {Biel}, \citenamefont {Niquet},\ and\ \citenamefont {Roche}}]{Lherbier2008}%
  \BibitemOpen
  \bibfield  {author} {\bibinfo {author} {\bibfnamefont {A.}~\bibnamefont {Lherbier}}, \bibinfo {author} {\bibfnamefont {B.}~\bibnamefont {Biel}}, \bibinfo {author} {\bibfnamefont {Y.-M.}\ \bibnamefont {Niquet}},\ and\ \bibinfo {author} {\bibfnamefont {S.}~\bibnamefont {Roche}},\ }\bibfield  {title} {\bibinfo {title} {Transport length scales in disordered graphene-based materials: Strong localization regimes and dimensionality effects},\ }\href {https://doi.org/10.1103/PhysRevLett.100.036803} {\bibfield  {journal} {\bibinfo  {journal} {Phys. Rev. Lett.}\ }\textbf {\bibinfo {volume} {100}},\ \bibinfo {pages} {036803} (\bibinfo {year} {2008})}\BibitemShut {NoStop}%
\end{thebibliography}

\begin{thebibliography}{5}%
\makeatletter
\providecommand \@ifxundefined [1]{%
 \@ifx{#1\undefined}
}%
\providecommand \@ifnum [1]{%
 \ifnum #1\expandafter \@firstoftwo
 \else \expandafter \@secondoftwo
 \fi
}%
\providecommand \@ifx [1]{%
 \ifx #1\expandafter \@firstoftwo
 \else \expandafter \@secondoftwo
 \fi
}%
\providecommand \natexlab [1]{#1}%
\providecommand \enquote  [1]{``#1''}%
\providecommand \bibnamefont  [1]{#1}%
\providecommand \bibfnamefont [1]{#1}%
\providecommand \citenamefont [1]{#1}%
\providecommand \href@noop [0]{\@secondoftwo}%
\providecommand \href [0]{\begingroup \@sanitize@url \@href}%
\providecommand \@href[1]{\@@startlink{#1}\@@href}%
\providecommand \@@href[1]{\endgroup#1\@@endlink}%
\providecommand \@sanitize@url [0]{\catcode `\\12\catcode `\$12\catcode `\&12\catcode `\#12\catcode `\^12\catcode `\_12\catcode `\%12\relax}%
\providecommand \@@startlink[1]{}%
\providecommand \@@endlink[0]{}%
\providecommand \url  [0]{\begingroup\@sanitize@url \@url }%
\providecommand \@url [1]{\endgroup\@href {#1}{\urlprefix }}%
\providecommand \urlprefix  [0]{URL }%
\providecommand \Eprint [0]{\href }%
\providecommand \doibase [0]{http://dx.doi.org/}%
\providecommand \selectlanguage [0]{\@gobble}%
\providecommand \bibinfo  [0]{\@secondoftwo}%
\providecommand \bibfield  [0]{\@secondoftwo}%
\providecommand \translation [1]{[#1]}%
\providecommand \BibitemOpen [0]{}%
\providecommand \bibitemStop [0]{}%
\providecommand \bibitemNoStop [0]{.\EOS\space}%
\providecommand \EOS [0]{\spacefactor3000\relax}%
\providecommand \BibitemShut  [1]{\csname bibitem#1\endcsname}%
\let\auto@bib@innerbib\@empty
\bibitem [{\citenamefont {Resta}(2018)}]{ModerntheoryinsulatingResta}%
  \BibitemOpen
  \bibfield  {author} {\bibinfo {author} {\bibfnamefont {R.}~\bibnamefont {Resta}},\ }\href {\doibase 10.1393/ncr/i2018-10151-1} {\bibfield  {journal} {\bibinfo  {journal} {Riv. Nuovo Cimento}\ }\textbf {\bibinfo {volume} {41}},\ \bibinfo {pages} {463} (\bibinfo {year} {2018})}\BibitemShut {NoStop}%
\bibitem [{\citenamefont {Lherbier}\ \emph {et~al.}(2008)\citenamefont {Lherbier}, \citenamefont {Biel}, \citenamefont {Niquet},\ and\ \citenamefont {Roche}}]{Lherbier2008}%
  \BibitemOpen
  \bibfield  {author} {\bibinfo {author} {\bibfnamefont {A.}~\bibnamefont {Lherbier}}, \bibinfo {author} {\bibfnamefont {B.}~\bibnamefont {Biel}}, \bibinfo {author} {\bibfnamefont {Y.-M.}\ \bibnamefont {Niquet}}, \ and\ \bibinfo {author} {\bibfnamefont {S.}~\bibnamefont {Roche}},\ }\href {\doibase 10.1103/PhysRevLett.100.036803} {\bibfield  {journal} {\bibinfo  {journal} {Phys. Rev. Lett.}\ }\textbf {\bibinfo {volume} {100}},\ \bibinfo {pages} {036803} (\bibinfo {year} {2008})}\BibitemShut {NoStop}%
\bibitem [{\citenamefont {Thonhauser}\ and\ \citenamefont {Vanderbilt}(2006)}]{haldanevanderblitlocalization}%
  \BibitemOpen
  \bibfield  {author} {\bibinfo {author} {\bibfnamefont {T.}~\bibnamefont {Thonhauser}}\ and\ \bibinfo {author} {\bibfnamefont {D.}~\bibnamefont {Vanderbilt}},\ }\href {\doibase 10.1103/PhysRevB.74.235111} {\bibfield  {journal} {\bibinfo  {journal} {Phys. Rev. B}\ }\textbf {\bibinfo {volume} {74}},\ \bibinfo {pages} {235111} (\bibinfo {year} {2006})}\BibitemShut {NoStop}%
\bibitem [{\citenamefont {Varjas}\ \emph {et~al.}(2020)\citenamefont {Varjas}, \citenamefont {Fruchart}, \citenamefont {Akhmerov},\ and\ \citenamefont {Perez-Piskunow}}]{computationtopomarker}%
  \BibitemOpen
  \bibfield  {author} {\bibinfo {author} {\bibfnamefont {D.}~\bibnamefont {Varjas}}, \bibinfo {author} {\bibfnamefont {M.}~\bibnamefont {Fruchart}}, \bibinfo {author} {\bibfnamefont {A.~R.}\ \bibnamefont {Akhmerov}}, \ and\ \bibinfo {author} {\bibfnamefont {P.~M.}\ \bibnamefont {Perez-Piskunow}},\ }\href {\doibase 10.1103/PhysRevResearch.2.013229} {\bibfield  {journal} {\bibinfo  {journal} {Phys. Rev. Res.}\ }\textbf {\bibinfo {volume} {2}},\ \bibinfo {pages} {013229} (\bibinfo {year} {2020})}\BibitemShut {NoStop}%
\bibitem [{\citenamefont {Alcón}\ \emph {et~al.}(2024)\citenamefont {Alcón}, \citenamefont {Cummings},\ and\ \citenamefont {Roche}}]{alcón2023tailoring}%
  \BibitemOpen
  \bibfield  {author} {\bibinfo {author} {\bibfnamefont {I.}~\bibnamefont {Alcón}}, \bibinfo {author} {\bibfnamefont {A.~W.}\ \bibnamefont {Cummings}}, \ and\ \bibinfo {author} {\bibfnamefont {S.}~\bibnamefont {Roche}},\ }\href {\doibase 10.1039/D3NH00416C} {\bibfield  {journal} {\bibinfo  {journal} {Nanoscale Horiz.}\ }\textbf {\bibinfo {volume} {9}},\ \bibinfo {pages} {407} (\bibinfo {year} {2024})}\BibitemShut {NoStop}%
\end{thebibliography}
\end{document}